\documentclass[twocolumn,showpacs,amsmath,amssymb,aps,pre,superscriptaddress,longbibliography]{revtex4-1}

\usepackage{graphicx}
\usepackage{dcolumn}
\usepackage{bm}
\usepackage{placeins}
\usepackage[english=american]{csquotes}
\usepackage[dvipsnames]{xcolor}

\begin{document}

\newcommand{\kb}{k_\text{B}}

\title{Gaussian white noise as a resource for microscopic engines}

\author{Andreas Dechant}
\affiliation{Department of Physics \#1, Graduate School of Science, Kyoto University, Kyoto 606-8502, Japan}

\author{Adrian Baule}
\affiliation{School of Mathematical Sciences, Queen Mary University of London, London E1 4NS, United Kingdom}

\author{Shin-ichi Sasa}
\affiliation{Department of Physics \#1, Graduate School of Science, Kyoto University, Kyoto 606-8502, Japan}

\begin{abstract}
We show that uncorrelated Gaussian noise, despite its paradigmatic association with thermal equilibrium, can drive a system out of equilibrium and can serve as a resource from which work can be extracted.
We consider an overdamped particle in a periodic potential with an internal degree of freedom and a state-dependent friction, coupled to an equilibrium bath.
Applying additional Gaussian white noise drives the system into a non-equilibrium steady state and causes a finite current if the potential is spatially asymmetric.
We calculate the current explicitly in the three complementary limits.
Since the particle current is driven solely by additive Gaussian white noise, this shows that the latter can potentially be exploited as a work resource to power small engines.
By comparing the extracted power to the energy injection due to the noise, we find an expression for the efficiency of such an engine.
\end{abstract}


\maketitle

\section{Introduction}
Noise, the random, uncontrollable and often unavoidable fluctuations in a system, is sometimes seen as merely a complication that one would rather be rid of.
Generally, however, this is far from true.
Noise can serve as a source of information about a system \cite{Lan98,Kob16}, or even be harnessed as a resource.
A particularly simple and well-studied for using noise as a resource is the class of systems referred to as Brownian ratchets \cite{Ast94,Doe94,Rou94,Jul97,Par98,Rei02,Han09}.
The latter are a prototypical model for microscopic engines, converting the random fluctuations due to noise into directed motion, effectively serving as noise rectifiers.
In accordance with the second law of thermodynamics, this rectification is not possible if the system is in thermal equilibrium, as was illustrated by Feynman \cite{Fey65}.
Ratchets thus necessarily have to be driven out of equilibrium to operate.
A variety of ratchets have been proposed and studied over the last two decades \cite{Jul97,Rei02,Han09} with applications in physics and biology.
They all share three crucial ingredients: (i) noise, (ii) broken symmetry (spatial or temporal) and (iii) nonequilibrium.
The paradigmatic model of a Brownian ratchet consists of an overdamped particle in a periodic potential and coupled to an equilibrium bath, represented by Gaussian white (uncorrelated) noise of intensity $D = \gamma T$, where $\gamma$ is the damping constant and $T$ is the temperature of the bath.
The symmetry breaking can be realized through a potential without inversion symmetry \cite{Ast94}, and the system can be driven out of equilibrium by a time-dependent change in the potential, e.~g.~by flashing the potential \cite{Rou94}.
Under these conditions, the particle can exhibit a net average drift, which can be used to perform work against an external load \cite{Kam98}.
Another way to drive the system out of equilibrium is to vary the temperature and thus noise intensity with time, either deterministically \cite{Rei96} or randomly \cite{Bao99}, which is akin to microscopic heat engines \cite{Sch07}.

Interestingly, directed motion can also be achieved with a static setup, i.~e.~with both potential and noise intensity being time-independent, if the noise distribution is non-Gaussian \cite{Luc95,Luc97,Cze97} or the noise is correlated in time (non-white) \cite{Mag93,Mil94,Doe94,Bar96}.
In these cases the term \enquote{noise rectifier} is even more appropriate, since no external driving is necessary but instead the noise itself drives the system out of equilibrium.
The ratchet thus directly converts the nonequilibrium fluctuations into a directed motion.
Noise rectification can also be realized in an underdamped system without potential, if the friction coefficient is a nonlinear function of the velocity \cite{Bau12,Bau13}. Here, in the absence of a potential the required asymmetry can be incorporated either in the friction nonlinearity or the noise itself. In all these examples of static noise rectifiers, the origin of the noise is non-thermal, i.~e.~, it represent the fluctuations of an environment that is not at thermal equilibrium. The noise thus essentially corresponds to a random force that injects power into the system, which, in turn, is converted into work.

We consider the case where the noise driving the system out of equilibrium is given as additive Gaussian white noise. 
This seems contradictory at first, since Gaussian white noise conventionally represents equilibrium fluctuations at a temperature $T = D/\gamma$, where $D$ is the noise intensity. 
This apparent uniqueness of Gaussian white noise raises the question of whether work can be extracted from Gaussian white noise at all. 

The clue to resolve this question comes from the observation that in underdamped systems with a velocity-dependent friction coefficient the addition of Gaussian white noise does not lead to a thermal equilibrium system since detailed balance is broken \cite{Dub09}. 
In this context, Gaussian white noise thus does not correspond to equilibrium fluctuations and can indeed be exploited to drive a current in a ratchet-type setting \cite{Sar13}. 
However, due to the inherent nonlinearity and coupled position and velocity degrees of freedom, this type of dynamics is not amenable to analytic treatment, in particular, there exists no general overdamped limit \cite{Hay05,Lin07}.

We present a simple overdamped model with similar properties as the general model of \cite{Sar13}, but which is amenable to analytical treatment. 
Our model, which we describe in detail in Section \ref{sec-model}, consists of an overdamped particle in an asymmetric potential, which is not only coupled to an equilibrium heat bath, but also posses an internal degree of freedom such that the friction coefficient depends on this internal state. 
If the switching between the internal states is induced by thermal activation, the system is at thermal equilibrium. 
However, if we now add Gaussian white noise, rectification occurs and the system exhibits a directed current. 
Our model permits an analytic treatment of the current in three distinct limiting regimes, which we analyze in detail in Section \ref{sec-limiting}: For switching dynamics of the internal state that is slow (Sec.~\ref{sec-slow}), respectively fast (Sec.~\ref{sec-fast}), compared to the diffusive dynamics in position space; and for the case of small noise magnitude $D$ (Sec.~\ref{sec-small-noise}).
We show that for slow switching the current is the same as for a system with a state-dependent temperature \cite{Bao99}.
Comparing the slow and fast switching regime (Sec.~\ref{sec-crossover}), we also predict the occurrence of current reversals.
Finally, we apply an external load to the ratchet to operate it as an engine and extract work.
Comparing the extracted work to the energy injected into the system by the added Gaussian white noise, we discuss the efficiency of the engine in Section \ref{sec-efficiency}.

\section{Model} \label{sec-model}
Our model consists of an overdamped particle moving in a periodic potential $U(x+L) = U(x)$ and immersed in an equilibrium bath at temperature $T$ with Stokes friction.
The motion of the particle is described by the overdamped Langevin equation
\begin{align}
m \gamma \dot{x} = - U'(x) + \sqrt{2 m \gamma T} \xi \label{eq-langevin} ,
\end{align}
where $x$ is the position of the particle, $m$ its mass (set to $1$ in the following) and $\xi$ denotes zero-mean Gaussian white noise of unit magnitude $\langle \xi(t) \xi(t') \rangle = \delta(t-t')$.
In addition, we assume that the particle has an internal degree of freedom, which can assume the discrete states $i = 1, \ldots, N$, and that the friction coefficient depends on this internal state $\gamma = \gamma_i$.
Physically this might be realized via a molecule that can have $N$ different conformations which differ in size.
Consequently, the Stokes friction coefficient, $\gamma_i = 6 \pi \eta R_i$, depends on the state $i$, with $R_i$ being the spatial extension of conformation $i$ and $\eta$ the viscosity of the medium.
Similar to the underdamped models with velocity dependent friction coefficient, the friction thus depends on an internal degree of freedom of the particle, which now, however, is discrete and independent the spatial motion.
We further define the rates $r_{i j} \geq 0$ at which the internal state changes from $j$ to $i$
\begin{align}
r_{i j} = r \alpha_{i j} \label{transition-rates}, 
\end{align}
where the $\alpha_{i j}$ are of order $1$.
Thus the parameter $r$ defines the typical timescale $\tau_r = 1/r$ on which the internal state changes.
We interpret the transitions between the internal states as a thermally activated process, i.~e.~the states are separated by energy barriers between which transitions occur due to thermal fluctuations at temperature $T$.
While this interpretation is not necessary from a mathematical point of view, it allows us to characterize the system as an equilibrium system at temperature $T$.
Equivalently to the Langevin equation \eqref{eq-langevin}, we can describe the system by a set of coupled Smoluchowski master equations for $i = 1, \ldots, N$,
\begin{align}
\partial_t P_{i,t} = \frac{1}{\gamma_i} \partial_x \Big[ U'(x) + T \partial_x \Big] P_{i,t} - \sum_{j} \big(r_{j i} P_i - r_{i j} P_{j,t} \big) \label{eq-fokkerplanck} ,
\end{align}
where $P_{i,t}(x) \text{d}x$ is the probability to find the particle at time $t$ in state $i$ and in the interval $[x,x+\text{d}x]$.
It is easily verified that the steady state solution to Eq.~\eqref{eq-fokkerplanck} for periodic boundary conditions is given by
\begin{subequations}
\begin{align}
&P_i(x) = \frac{\mathcal{P}_i}{Z} e^{-\frac{U(x)}{T}}, \label{eq-solution-position} \\
&\sum_{j} \alpha_{j i} \mathcal{P}_i = \sum_{j} \alpha_{i j} \mathcal{P}_j, \label{eq-solution-states}  \\
&P(x) = \sum_i P_i(x) = \frac{\sum_i \mathcal{P}_i }{Z} e^{-\frac{U(x)}{T}} = \frac{1}{Z} e^{-\frac{U(x)}{T}} \label{eq-solution-BG} .
\end{align} \label{eq-solution}%
\end{subequations}%
Here $Z = \int_0^L \text{d}x \ e^{-U(x)/T}$ is the partition function and the $\mathcal{P}_i$ are the occupation probabilities in state $i$, with the normalization condition $\sum_i \mathcal{P}_i = 1$.
The occupation probabilities are determined by the solution to Eq.~\eqref{eq-solution-states} and normalization.
In the stationary state the position and the internal state decouple and the position distribution is precisely the equilibrium Boltzmann-Gibbs distribution Eq.~\eqref{eq-solution-BG}.
The system thus reaches an equilibrium state at temperature $T$ and there is no net motion in the stationary state.

To drive the system out of equilibrium, we introduce an additional Gaussian white noise with constant intensity $D$,
\begin{align}
\gamma_i \dot{x} = - U'(x) + \sqrt{2 \gamma_i T} \xi + \sqrt{2 D} \eta \label{langevin} ,
\end{align}
with $\langle \eta(t) \eta(t') \rangle = \delta(t-t')$ and $\langle \eta(t) \xi(t') \rangle = 0$.
The steady state Smoluchowski master equation is then modified to
\begin{align}
\frac{1}{\gamma_i} \partial_x \Big[ U'(x) + T_i \partial_x \Big] P_i = r \sum_{j} \big(\alpha_{j i} P_i - \alpha_{i j} P_j \big) \label{fokkerplanck} ,
\end{align}
where we defined the state-dependent effective temperature $T_i = T + D/\gamma_i$.
This obviously reduces to Eq.~\eqref{eq-fokkerplanck} for $D=0$.
For finite noise intensity $D$, however, the solution is no longer given by Eq.~\eqref{eq-solution}; due to the state-dependent temperature, the position and internal state are now coupled.
We note that the steady state of the occupation probabilities $\mathcal{P}_i = \int_0^L \text{d}x \ P_i(x)$ still has to satisfy Eq.~\eqref{eq-solution-states}.
Formally, Eq.~\eqref{fokkerplanck} is similar to the ratchet model introduced in Ref.~\cite{Bao99} with a fluctuating temperature.
However, the physical picture is quite different, as in our model the effective temperatures $T_i$ are arise through the combination of state-dependent friction and Gaussian white noise with \textit{constant} intensity.
By contrast, for the model of Ref.~\cite{Bao99}, the friction is constant $\gamma_i \equiv \gamma$ and the $T_i$ correspond to different physical temperatures.
In the following analysis, we first treat $T_i$ and $\gamma_i$ as independent, as this allows for the results to be applied either situation.

The physical origin of the additional Gaussian white noise can be understood in the following way:
Consider an additional heat bath at temperature $T^*$, whose coupling $\gamma^*$ to the system is independent of the internal state.
Formally this then leads to a modified friction coefficient $\tilde{\gamma}_i = \gamma_i + \gamma^*$ and effective temperature, $\tilde{T}_i = \gamma_i/\tilde{\gamma}_i T + \gamma^*/\tilde{\gamma}_i T^*$.
It is then easy to see that the system is out of equilibrium (i.~e.~$\tilde{T}_i \neq \tilde{T}_j$) for $T^* \neq T$.
In particular for $T^{*} \rightarrow \infty$ and $\gamma^* \rightarrow 0$ but $D = \gamma^* T^*$ finite, one obtains precisely the model with added Gaussian white noise Eq.~\eqref{langevin} introduced above.
Thus additional Gaussian white noise corresponds to a second, high-temperature but weakly coupled heat bath.
A concrete experimental realization could be a charged particle, which changes between two conformations of different size and thus friction coefficients $\gamma_{1}$ and $\gamma_2$, embedded in a viscous fluid at temperature $T$.
The additional white noise may realized as an electrostatic force due to charge fluctuations on a capacitor, either due to thermal Johnson-Nyquist noise at temperature $T^* \gg T$ or an artificially generated random voltage applied to the capacitor.
Imposing an asymmetric periodic potential should then lead to an observable current.

\section{Limiting cases} \label{sec-limiting} 
Our model is simple enough that it permits an analytic treatment, albeit not in full generality.
In the following four sections, we will thus focus on three limiting cases, where the transition rate between the internal states is slow (Sec.~\ref{sec-slow}) or fast (Sec.~\ref{sec-fast}) compared to the dynamics in the periodic potential and where the intensity of the external noise is small (Sec.~\ref{sec-small-noise}).
In all three limits, we show that the system exhibits a non-vanishing current for nonzero $D$ and thus work can be extracted from the added Gaussian white noise.
To facilitate the discussion of the three limits, we introduce an overall damping strength $\gamma$ and temperature $T$, which are related to the friction coefficient and effective temperatures in the individual states by $\gamma_i = \nu_i \gamma$ and $T_i = \theta_i T$, where the $\nu_i$ and $\theta_i$ are dimensionless quantities that we assume to be of order $1$.
Rescaling the position coordinate by the period $L$ of the potential $x = z L$ with $0 \leq z \leq 1$, we can write Eq.~\eqref{fokkerplanck} in dimensionless form
\begin{align}
\frac{1}{\nu_i} \partial_z \Big[u'(z) &+ \theta_i \partial_z \Big] p_i(z) = \rho \sum_{j} \big(\alpha_{j i} p_i - \alpha_{i j} p_j \big) \label{fokkerplanck-dimensionless} .
\end{align}
Here $u(z) = U(L z)/T$ is a rescaled potential with period $1$ and $p_i(z) = L P_i(L z)$.
For not too deep potentials $u \sim O(1)$, the parameter $\rho = r \gamma L^2/T = \tau_x/\tau_r$ specifies the relaxation time $\tau_x = \gamma L^2/T $ of the dynamics along the spatial direction relative to the one internal states, $\tau_r = 1/r$.
For $\rho \ll 1$ ($\tau_r \gg \tau_x$) the spatial dynamics are much faster than the transitions between the internal states, which we refer to as slow switching.
Conversely, we refer to the opposite case $\rho \gg 1$ ($\tau_r \ll \tau_x$) as fast switching.
Writing $\theta_i = 1 + \mathcal{D}/\nu_i$, the relative magnitude of the external noise is quantified by the parameter $\mathcal{D} = D/(\gamma T)$, with $\mathcal{D} \ll 1$ corresponding to weak external noise.

\subsection{Slow switching} \label{sec-slow}
We start with discussing the slow switching regime, where the internal state changes slowly compared to the position dynamics.
We expand the solution to Eq.~\eqref{fokkerplanck-dimensionless} in terms of parameter $\rho$,
\begin{align}
p_i(x) \simeq p_i^{(0)}(z) + \rho p_i^{(1)}(z) + O(\rho^2) \label{small-r-expansion} .
\end{align}
To lowest order (i.~e.~for $\rho \rightarrow 0$), the solution to Eq.~\eqref{fokkerplanck-dimensionless} is readily obtained,
\begin{align}
p_i^{(0)}(z) = \frac{\mathcal{P}_i}{Z_i} e^{-\frac{u(z)}{\theta_i}} \label{r0-solution} ,
\end{align}
with $Z_i = \int_0^1 \text{d}z \ e^{-\frac{u(z)}{\theta_i}}$.
In this limit, the particle is with probability $\mathcal{P}_i$ in a quasi-equilibrium state at effective temperature $T_i$.
Note that the occupation probabilities $\mathcal{P}_i$ are still assumed to be stationary; this requires that the stationary limit $t \rightarrow \infty$ is taken before taking $r \rightarrow 0$.
The quasi-equilibrium nature of the steady state means that to zeroth order in $\rho$, no current can be observed.
To first order in $\rho$ we find a set of equations for the sub-leading order corrections,
\begin{align}
\frac{1}{\nu_i} \partial_z \Big[ u'(z) + \theta_i \partial_z \Big] p_i^{(1)} = \sum_{j} \big(\alpha_{j i} p^{(0)}_i - \alpha_{i j} p^{(0)}_j \big) \label{fokkerplanck-order1} .
\end{align}
The general solution to Eq.~\eqref{fokkerplanck-order1} reads
\begin{align}
p^{(1)}_i(z) &= \frac{\nu_i}{\theta_i} e^{-\frac{u(z)}{\theta_i}} \Big[ \mathcal{N}^1_i + \int_0^z \text{d}y \ e^{\frac{u(y)}{\theta_i}} \Big( f_i(y) - \mathcal{J}^1_i \Big) \Big] \label{r1-solution}\\
\text{with} \quad &f_i(y) = \sum_{j} \int_0^y \text{d}x \ \big(\alpha_{j i} p^{(0)}_i(x) - \alpha_{i j} p^{(0)}_j(x) \big) . \nonumber
\end{align}
The constants $\mathcal{N}^1_i$ and $\mathcal{J}^1_i$ have to be determined from the normalization $\int_0^1 \text{d}z \ p_i^1(z) = 0$ \cite{note1}
and periodicity $p_i^1(z+1) = p_i^1(z)$ conditions.
We obtain
\begin{subequations}
\begin{align}
\mathcal{J}_i^1 &= \frac{1}{Z_i^+} \int_0^1 \text{d}z \ e^{\frac{u(z)}{\theta_i}} f_i(z) , \\
\mathcal{N}_i^1 &= \frac{1}{Z_i} \int_0^1 \text{d}z \int_0^z \text{d}y \ e^{-\frac{u(z)-u(y)}{\theta_i}} \big( \mathcal{J}_i^1 - f_i(y) \big) ,
\end{align} \label{r1-solution-constants}%
\end{subequations}
where we defined $Z_i^+ = \int_0^1 \text{d}x \ e^{u(z)/\theta_i}$.
The constants $\mathcal{J}_i^1$ can be related to the total probability current.
Summing Eq.~\eqref{fokkerplanck-dimensionless} over $i$, we obtain
\begin{align}
\partial_z \sum_i \frac{1}{\gamma_i} \Big[ u'(z) + \theta_i \partial_z \Big] p_i(z) \equiv - \partial_z \sum_i J_i(z) = 0 \label{current-def} ,
\end{align}
since the double sum on the right hand side of Eq.~\eqref{fokkerplanck} cancels.
The total probability current $J$ is defined via $\tau_x \partial_t p_t(z) = - \partial_z J_z$ and in the stationary state we have $\partial_z J_z = 0$, so that $J_z \equiv J$ independent of $z$.
Comparing this to Eq.~\eqref{fokkerplanck-dimensionless}, we see that $J_z = \sum_i J_i(z)$.
Expanding $J_i$ in terms of $\rho$, we have $J_i(x) = J_i^{(0)}(z) + \rho J_i^{(1)}(z)$.
From Eq.~\eqref{r1-solution} we identify $\mathcal{J}_i^1 = J_i^{(1)}(0)$.
Since the zeroth order currents vanish and $J$ is independent of $z$, we thus have $J = \rho \sum_i \mathcal{J}_i^1$.
The dimensionless current is then given by
\begin{widetext}
\begin{align}
J_z = \rho \sum_{i, j} \frac{1}{Z_i^+} \int_0^1 \text{d}z \ e^{\frac{u(z)}{\theta_i}} \int_0^z \text{d}y \ \bigg(\frac{\alpha_{j i} \mathcal{P}_i}{Z_i} e^{-\frac{u(y)}{\theta_i}} - \frac{\alpha_{i j} \mathcal{P}_j}{Z_j} e^{-\frac{u(y)}{\theta_j}} \bigg) . \label{current-slow-result}
\end{align}
\end{widetext}
It is related to the physical current $J_x$ via $J_x = J_z/\tau_x$
Equation.~\eqref{current-slow-result} constitutes the first main result of our paper.
For nonzero $\mathcal{D}$, state-dependent friction and an asymmetric potential, the system exhibits a finite current and drift velocity.
We note that $J_x$ depends on the friction coefficients $\gamma_i$ only via the effective temperatures $T_i$.
In the slow switching regime, the current obtained is equivalent to a system switching between physical temperatures $T_i$ as in the model of Ref.~\citep{Bao99}.
The physical reason for this is that for $r \ll T/(\gamma L^2)$, the system in state $i$ is most of the time close to the quasi-equilibrium state at effective temperature $T_i$, which depends on the damping coefficient $\gamma_i$ only via the effective temperature $T_i$.

\begin{figure}[ht!]
\includegraphics[width=0.23\textwidth]{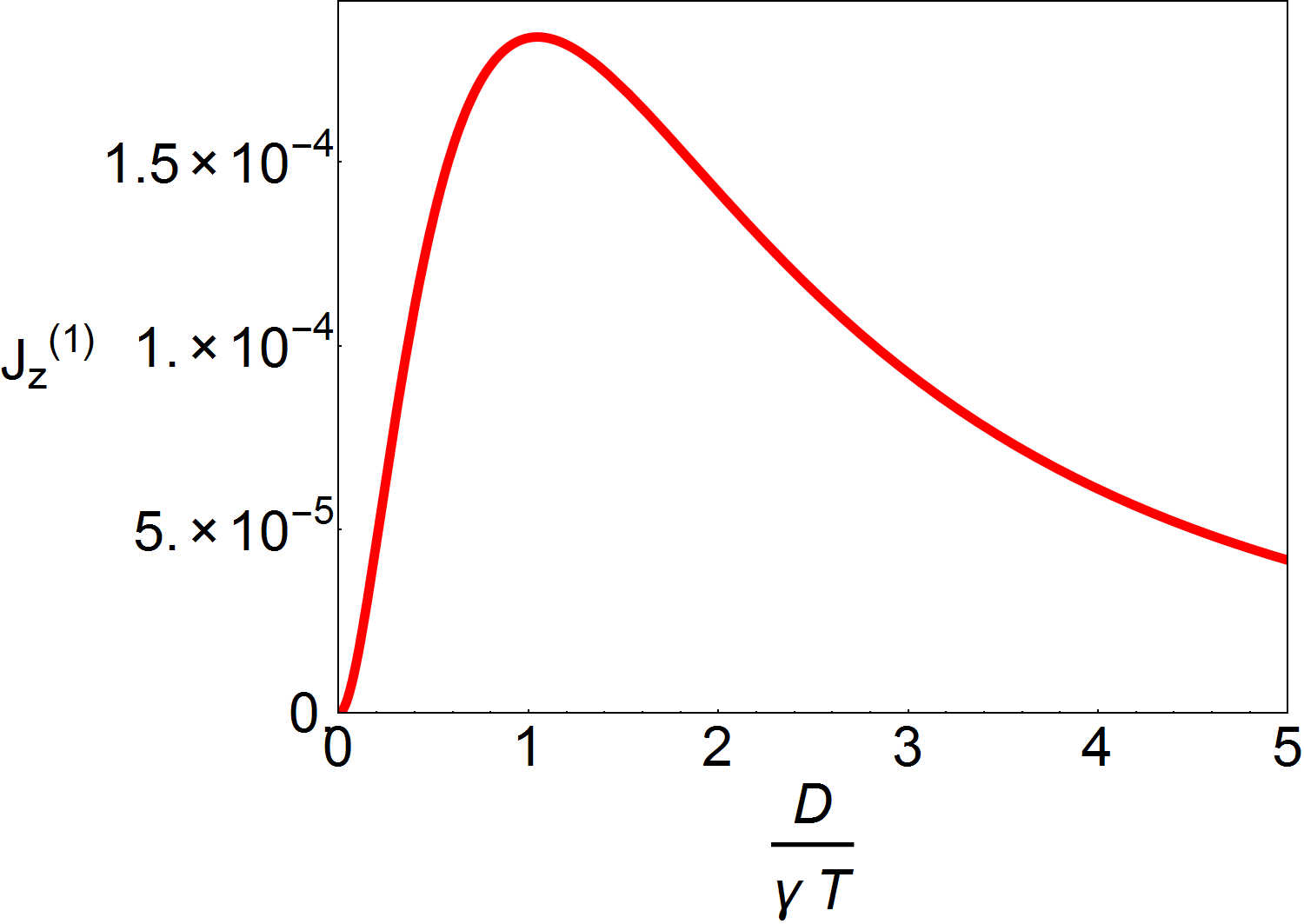}
\includegraphics[width=0.23\textwidth]{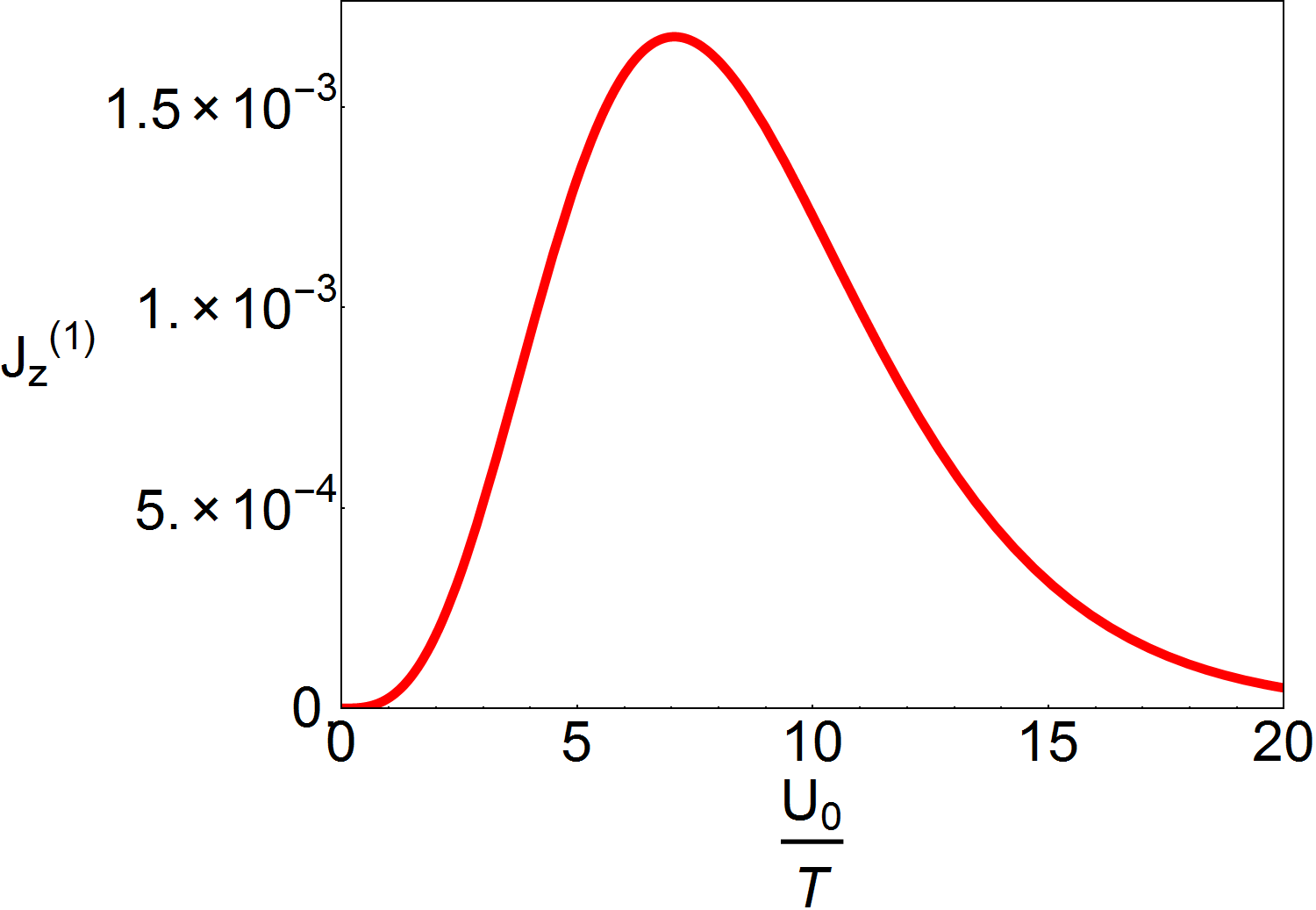}\\
\includegraphics[width=0.23\textwidth]{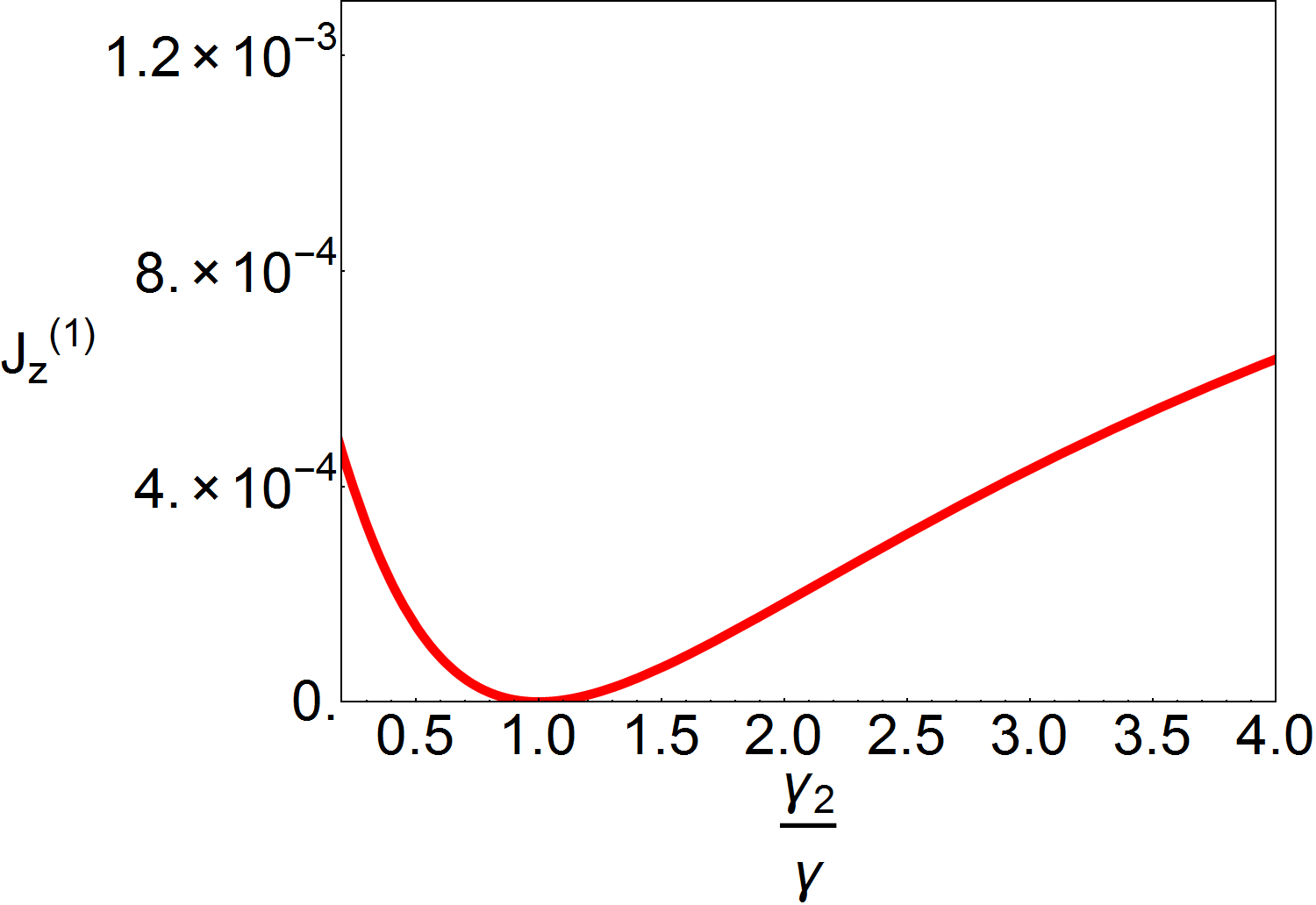}
\includegraphics[width=0.23\textwidth]{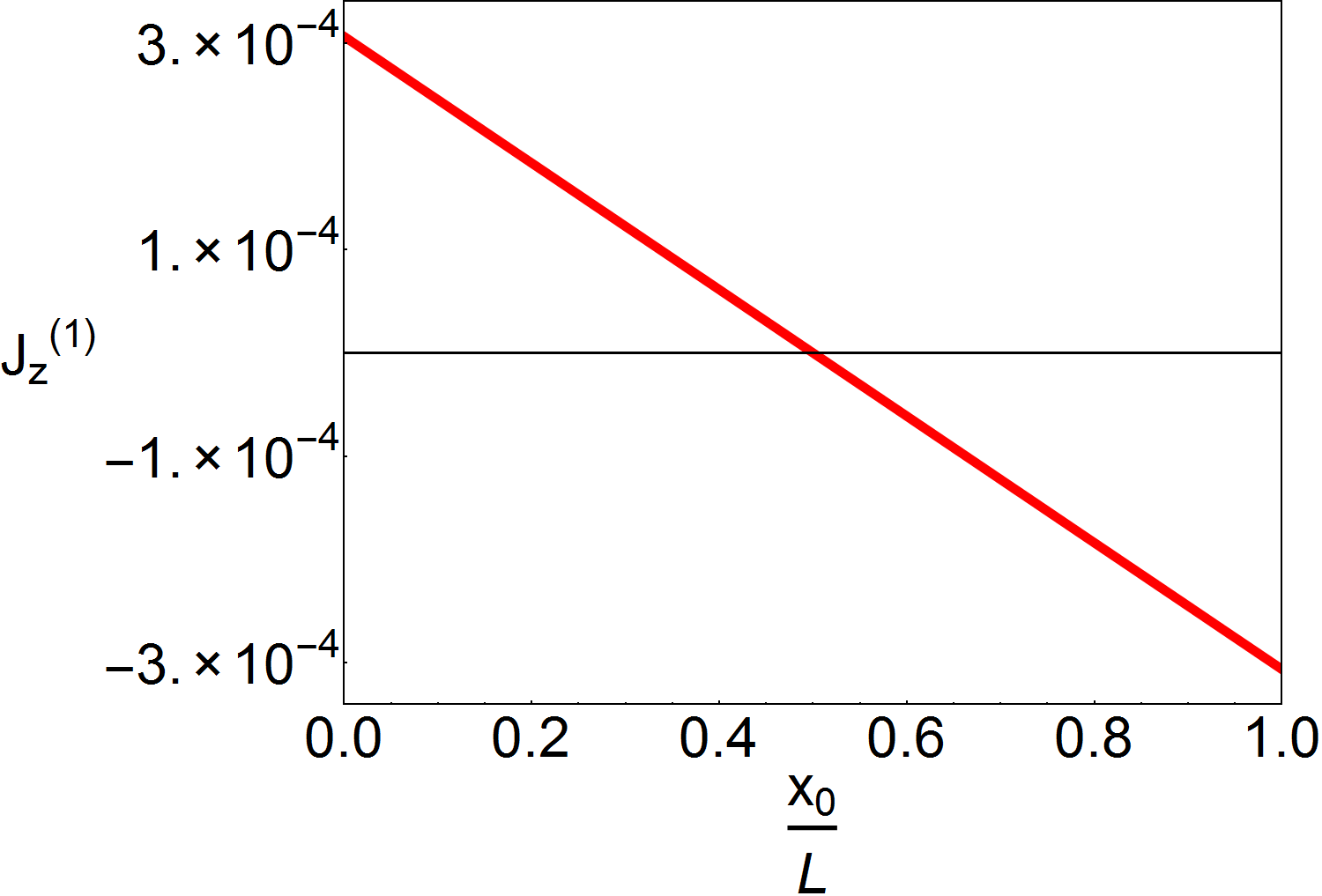}
\caption{The first order contribution to the dimensionless current $J_z^{(1)}$ for slow switching, Eq.~\eqref{current-slow-result} as a function of the noise intensity $D$ (top left), potential depth $U_0$ (top right), friction coefficient $\gamma_2$ (bottom left) and asymmetry parameter $x_0$ (bottom right). The respective remaining parameters are: $D = \gamma T, \ \gamma_2 = 2 \gamma, \ \gamma_1 = \gamma, \ U_0 = 2 T, \ x_0 = 0.2 L$. Note that for $D \rightarrow 0$, the current approaches $0$ as $D^2$. \label{fig:current-slow} }
\end{figure}

As a specific example, we choose a piecewise linear potential
\begin{align}
U(x) = U_0 \times \left\lbrace \begin{array}{ll}
\frac{x}{x_0} &\text{for} \; 0 \leq x < x_0 \\[2 ex]
\frac{L-x}{L-x_0} &\text{for} \; x_0 \leq x \leq L ,
\end{array} \right. \label{piecewise-potential}
\end{align}
for which the integrals in Eq.~\eqref{current-slow-result} can be evaluated explicitly.
The result for two internal states $i = 1,2$ is given in Eq.~\eqref{current-slow-explicit} and is shown graphically in Fig.~\ref{fig:current-slow} as a function of different parameters.
The current vanishes for a symmetric potential $x_0 = L/2$, identical friction coefficients $\gamma_1 = \gamma_2$ or zero external noise $D = 0$.
The current is generally larger, the more asymmetric the potential or the more pronounced the difference in friction coefficients and thus effective temperatures is.
As a function of $D$, the current shows a $D^2$ behavior for small $D$; it then reaches a maximum at intermediate $D$ before decreasing again at large $D$.
In Sec.~\ref{sec-small-noise} we show that the $D^2$ scaling for small $D$ is in fact generic and not limited to the slow switching regime.
For large $D$, the overall effective temperature is increased and thus the influence of the potential, whose asymmetry generates the current, becomes negligible.
The current also has a maximum at intermediate values of the potential depth $U_0$.
For small $U_0$ compared to $T$, the dynamics are mostly diffusive, whereas for large $U_0$ the particle will be trapped near the minimum of the potential; in both situations the current is small.

\subsection{Fast switching} \label{sec-fast}
\begin{figure}[ht!]
\includegraphics[width=0.23\textwidth]{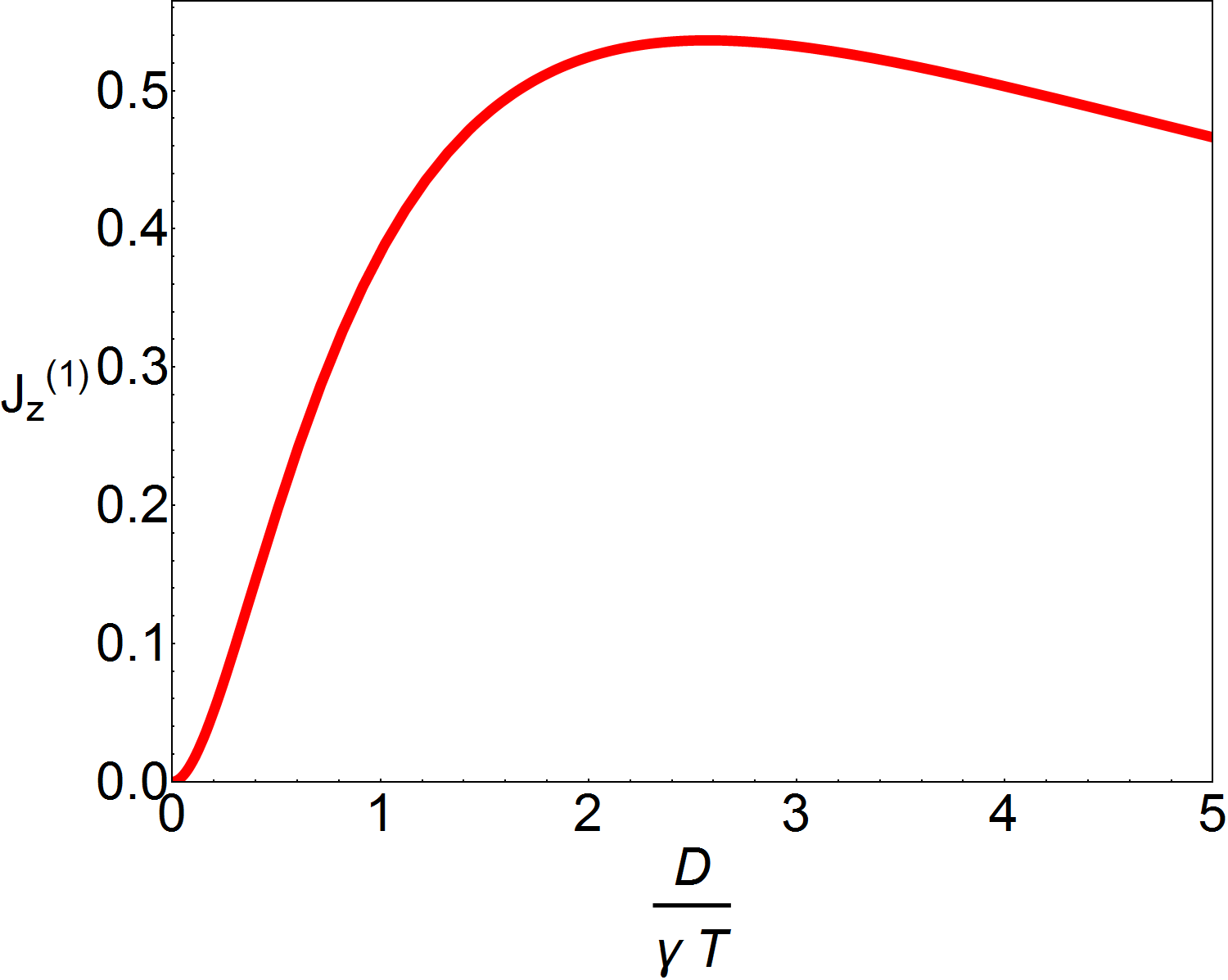}
\includegraphics[width=0.23\textwidth]{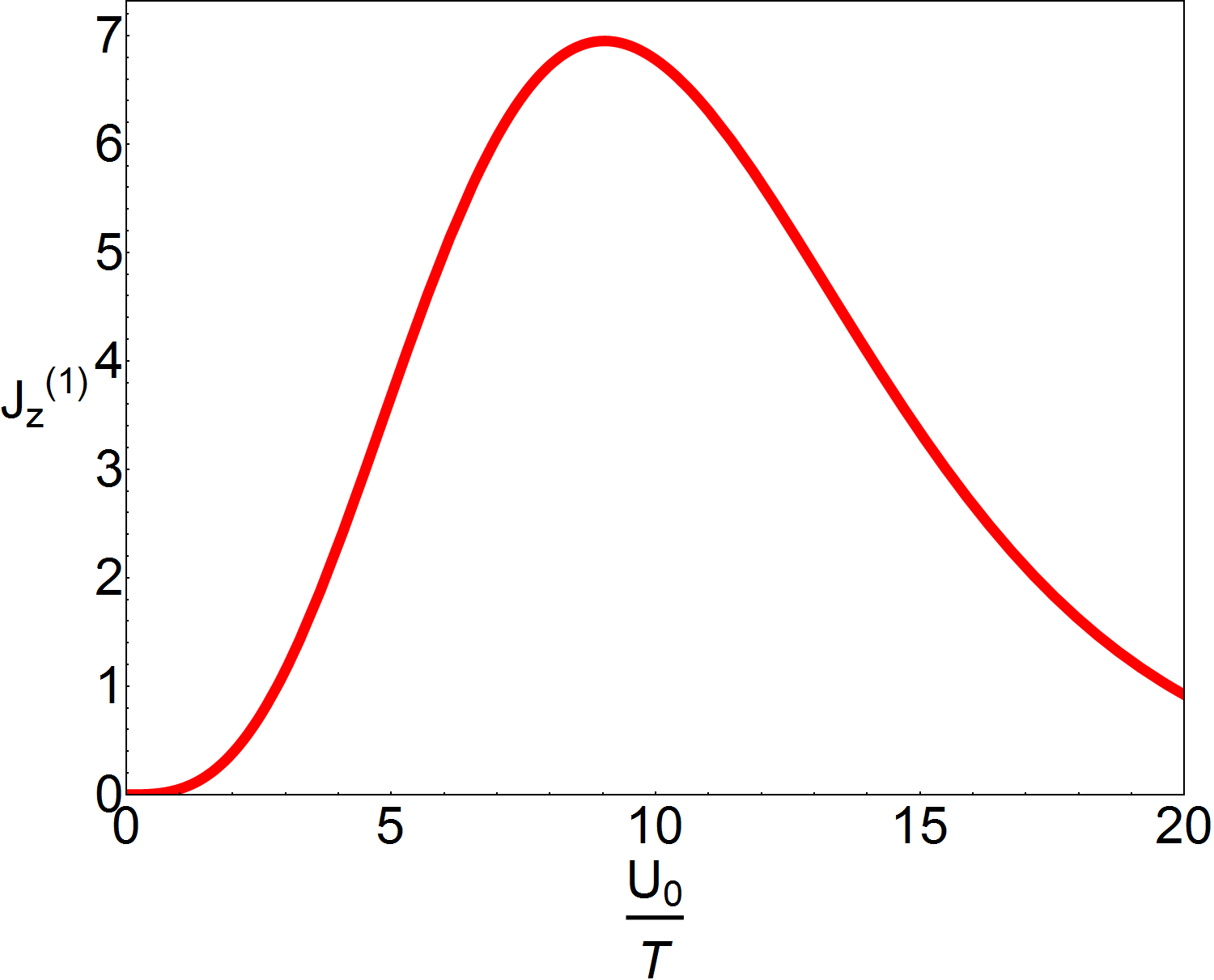}\\
\includegraphics[width=0.23\textwidth]{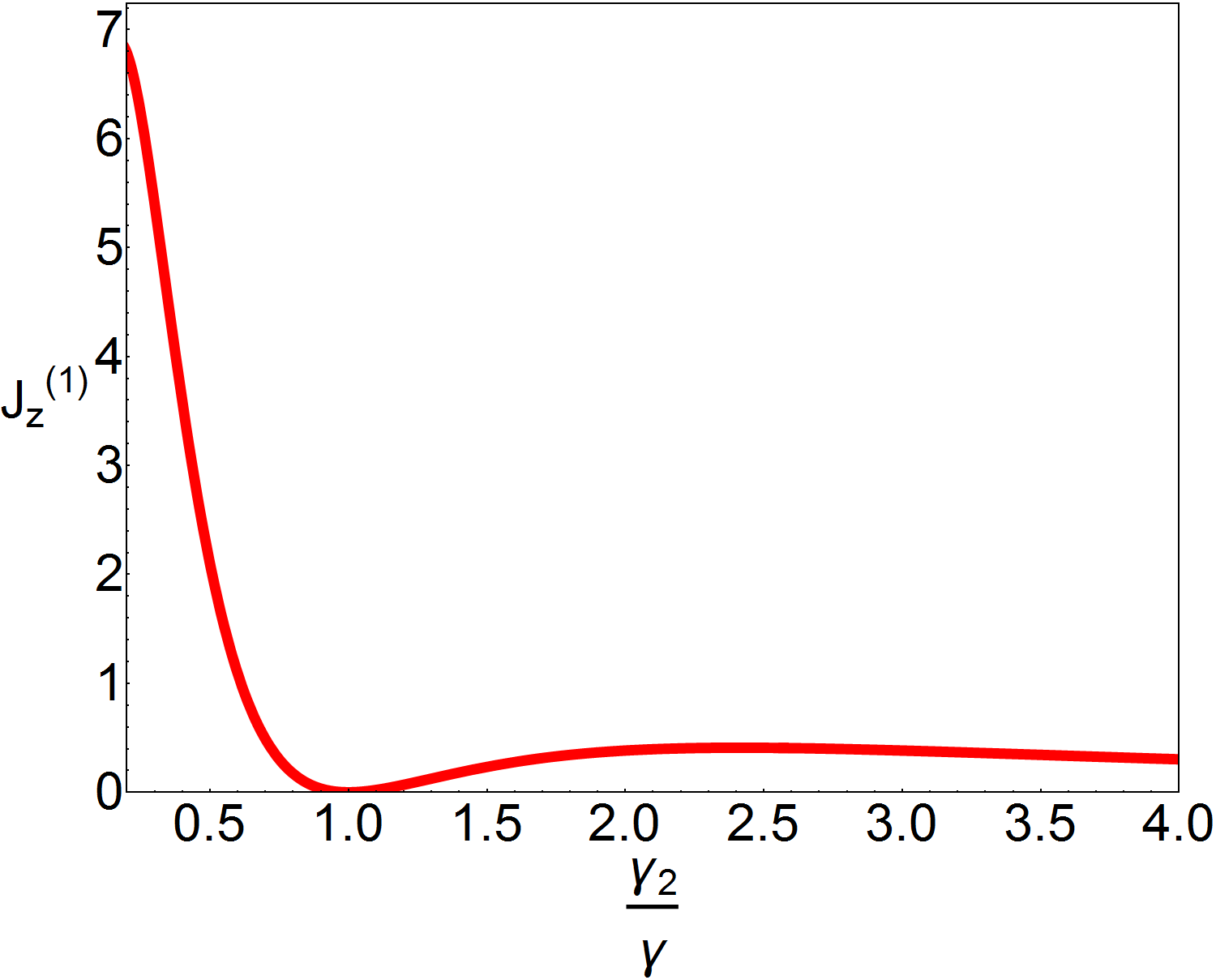}
\includegraphics[width=0.23\textwidth]{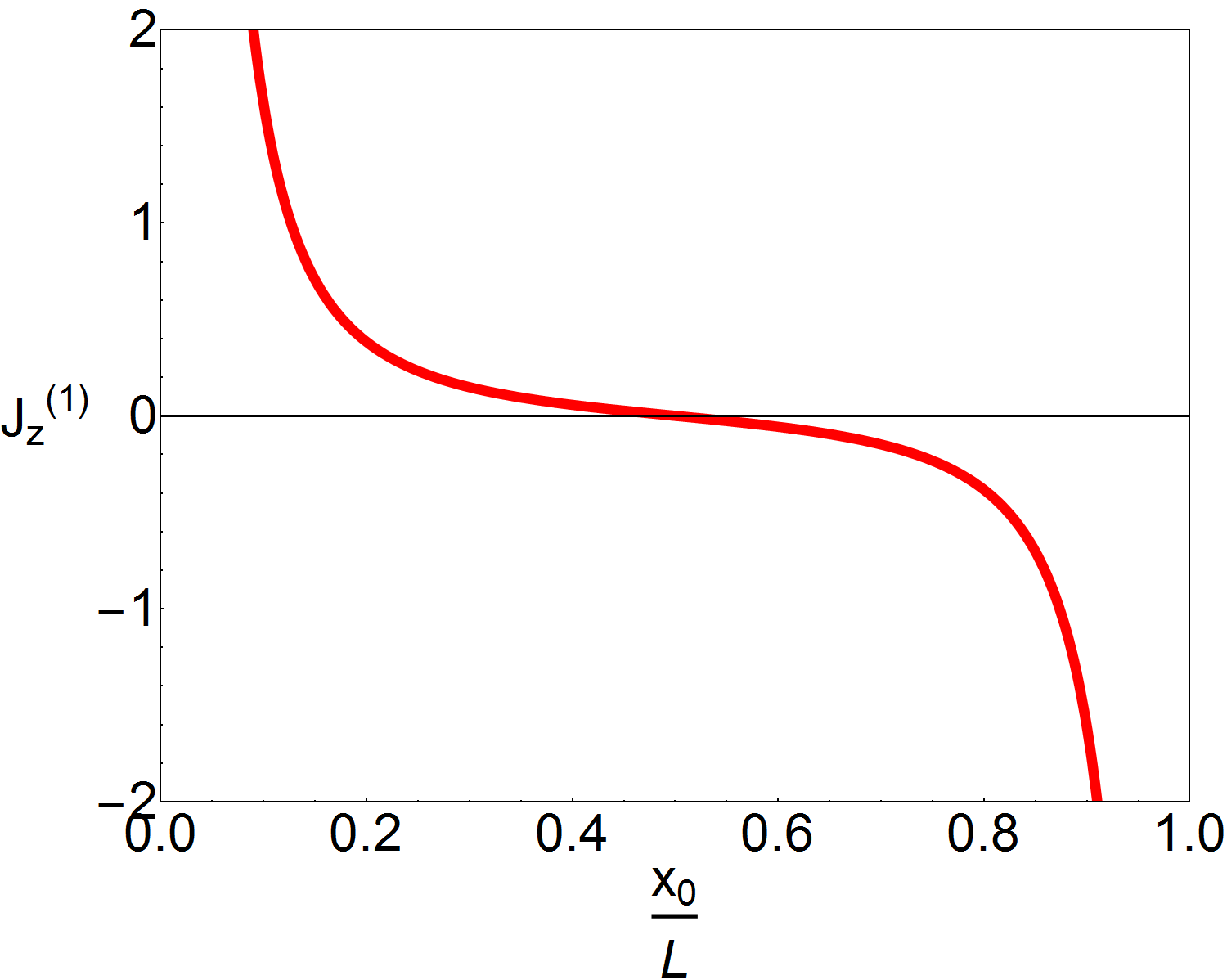}
\caption{The first order contribution to the dimensionless current $J_z^{(1)}$ for fast switching, Eq.~\eqref{current-fast-result} as a function of the noise intensity $D$ (top left), potential depth $U_0$ (top right), friction coefficient $\gamma_2$ (bottom left) and asymmetry parameter $x_0$ (bottom right). The respective remaining parameters are: $D = \gamma T, \ \gamma_2 = 2 \gamma, \ \gamma_1 = \gamma, \ U_0 = 2 T, \ x_0 = 0.2 L$. \label{fig:current-fast} }
\end{figure}

In the opposite limit, the transitions between the internal states are much faster than the dynamics along the coordinate axis, $\rho \gg 1$ or $r \gg T/(\gamma L^2)$.
For fluctuating physical temperature, this was examined in Ref.~\citep{Bao99}, and the leading order contribution to the current was found to be proportional to the inverse switching rate $1/r$.
We now re-examine this limit for our model, with state-dependent friction.
We restrict ourselves to two internal states with transition rates $r_{1 2} = r_{2 1} = r$.
Expanding
\begin{align}
p_i(z) \simeq p_i^{(0)}(z) + \rho^{-1} p_i^{(1)}(z) + O(\rho^{-2}) \label{large-r-expansion} ,
\end{align}
we find from Eq.~\eqref{fokkerplanck-dimensionless} to order $\rho^{-1}$
\begin{subequations}
\begin{align}
&p_1^{(0)}(z) - p_2^{(0)}(z) = 0 \label{p0-equation} \\
&p_1^{(1)}(z) - p_2^{(1)}(z) = \frac{1}{\gamma_1} \partial_z \Big[ u'(z) + \theta_1 \partial_z \Big] p_1^{(0)}(z) \label{p1-equation-1} \\
&p_2^{(1)}(z) - p_1^{(1)}(z) = \frac{1}{\gamma_2} \partial_z \Big[ u'(z) + \theta_2 \partial_z \Big] p_2^{(0)}(z) \label{p1-equation-2}.
\end{align}%
\end{subequations}
Since Eq.~\eqref{p0-equation} has to be true at any $z$, it can only be satisfied for
\begin{align}
p_i^{(0)}(z) = \mathcal{P}_i \hat{p}^{(0)}(z),
\end{align}
with $\mathcal{P}_1 = \mathcal{P}_2 = 1/2$ where $\hat{p}^{(0)}$ is independent of $i$.
Summing Eqs.~\eqref{p1-equation-1} and \eqref{p1-equation-2} we then obtain an equation for $\hat{p}^{(0)}$,
\begin{align}
&\frac{1}{\overline{\nu}} \partial_z \Big[u'(z) + \overline{\theta} \partial_z \Big] \hat{p}^{0}(z) , \\
\text{with} \quad &\frac{1}{\overline{\nu}} = \frac{1}{\nu_1} + \frac{1}{\nu_2}, \quad \overline{\theta} = \overline{\nu} \bigg( \frac{\theta_1}{\nu_1} + \frac{\theta_2}{\nu_2} \bigg) \nonumber .
\end{align}
The zeroth-order solution is then just the Boltzmann-Gibbs density at effective temperature $\overline{\theta}$,
\begin{align}
p_i^{(0)}(z) = \frac{\mathcal{P}_i}{\overline{Z}} e^{-\frac{u(z)}{\overline{\theta}}} \label{p0-solution} ,
\end{align}
with $\overline{Z} = \int_0^1 \text{d}z \ e^{-u(z)/\overline{\theta}}$.
As expected, the zeroth-order contribution to the current vanishes also in the fast switching regime.
Adding the order $\rho^{-2}$ equations analog to Eqs.~\eqref{p1-equation-1} and \eqref{p1-equation-2}, we obtain an equation for $p_i^{(1)}$,
\begin{align}
\frac{1}{\nu_1} \partial_z &\Big[ u'(x) + \theta_1 \partial_z \Big] p_1^{(1)}(z) \\
&+\frac{1}{\nu_2} \partial_z \Big[ u'(z) + \theta_2 \partial_z \Big] p_2^{(1)}(z) = 0 \nonumber .
\end{align}
Recalling the definition of the current, Eq.~\eqref{current-def}, we can equivalently write this as
\begin{align}
\frac{1}{\nu_1} &\Big[ u'(z) + \theta_1 \partial_z \Big] p_1^{(1)}(z) \\
&+\frac{1}{\nu_2} \Big[ u'(z) + \theta_2 \partial_z \Big] p_2^{(1)}(z) = -J_z^{(1)} \nonumber .
\end{align}
We now use Eq.~\eqref{p1-equation-2} to eliminate $p_2^{(1)}$,
\begin{align}
&\frac{1}{\overline{\nu}} \Big[ u'(z) + \overline{\theta} \partial_z \Big] p_1^{(1)}(z) = -J_z^{(1)} \label{p1-solution} \\
& - \frac{1}{\nu_2^2} \Big[ u'(z) + \theta_2 \partial_z \Big] \partial_z \Big[ u'(z) + \theta_2 \partial_z \Big] p_2^{(0)}(z)  \nonumber .
\end{align}
Solving for $p_1^{(1)}$ and imposing periodicity, $p_1^{(1)}(z+1)=p_1^{(1)}(z)$, we find for the first order current $J_z^{(1)}$,
\begin{align}
J_z^{(1)} &= -\frac{1}{\nu_2^2 \overline{Z}^+} \int_0^1 \text{d}z \ e^{\frac{u(z)}{\overline{\theta}}} \\
&\quad \times \Big[ u'(z) + \theta_2 \partial_z \Big] \partial_z \Big[ u'(z) + \theta_2 \partial_z \Big] p_2^{(0)}(z) \nonumber,
\end{align} 
where we defined $\overline{Z}^+ = \int_0^1 \text{d}z \ e^{u(z)/\overline{\theta}}$.
Plugging in the result for $p_2^{(0)}$ and integrating by parts, we obtain for the total current,
\begin{align}
J_z &= -\frac{1}{\rho} \frac{\overline{\nu}^2(\theta_1 - \theta_2)^2}{2(\nu_1 \nu_2)^2 \overline{Z} \overline{Z}^+ \overline{\theta}^3} \int_{0}^{1} \text{d}z \ \big(u'(z)\big)^3 \label{current-fast-result} ,
\end{align}
which, as before, is related to the physical current via $J_x = J_z/\tau_x$.
This is our second main result.
We note that for $\gamma_1 = \gamma_2$ this reduces to the result of Ref.~\citep{Bao99}, provided that $T_1$ and $T_2$ are interpreted as physical temperatures.
For the model with state-dependent friction, we have effective temperatures with $T_1 - T_2 = D (1/\gamma_1 - 1/\gamma_2)$.
It is then straightforward to see that, just as in the fast switching case, the current is proportional to $D^2$ for small $D$.
Eq.~\eqref{current-fast-result} is easily evaluated for the piecewise linear potential Eq.~\eqref{piecewise-potential} and reads, after returning to dimensionful quantities,
\begin{align}
J_x &= -\frac{D^2}{r} \bigg(\frac{\gamma_1 - \gamma_2}{\gamma_1 \gamma_2 (\gamma_1 + \gamma_2)} \bigg)^2 \bigg( \frac{U_0}{\overline{T}} \bigg)^5 \frac{\frac{1}{2} - \frac{x_0}{L}}{L^4 \big(\frac{x_0}{L} \big( 1 - \frac{x_0}{L}\big) \big)^2}  \nonumber \\
&\qquad \qquad \times \Big(4 \cosh\Big(\frac{U_0}{\overline{T}}\Big) - 4 \Big)^{-1}. \label{current-fast-explicit}
\end{align}
The current for fast switching in the piecewise linear potential is shown in Fig.~\ref{fig:current-fast} as a function of various parameters.
The overall dependence on the parameters is similar to the slow switching case; in particular the current also vanishes for a symmetric potential $x_0 = L/2$, identical friction coefficients $\gamma_1 = \gamma_2$ or zero external noise $D = 0$.
Note that contrary to the slow switching regime, the current diverges as $x_0$ approaches $0$ or $L$, which corresponds to a jump in the potential.
In this situation, the fast switching approximation breaks down, since at the discontinuity, the spatial motion of the particle can no longer be described by the slow timescale $\tau_x$.
The most notable difference to the slow switching results is that the magnitude of the first order correction is considerably larger.
We discuss the consequence of this finding in the next Section.
In contrast to the slow switching result Eq.~\eqref{current-slow-result}, for fast switching the current Eq.~\eqref{current-fast-result} explicitly depends on the friction coefficients $\gamma_i$ instead of just the effective temperatures $T_i$.
For fast switching, the current is thus different for state dependent friction as compared to a fluctuating temperature \citep{Bao99}.

\subsection{Crossover between slow and fast switching} \label{sec-crossover}
\begin{figure}[ht!]
\includegraphics[width=0.23\textwidth]{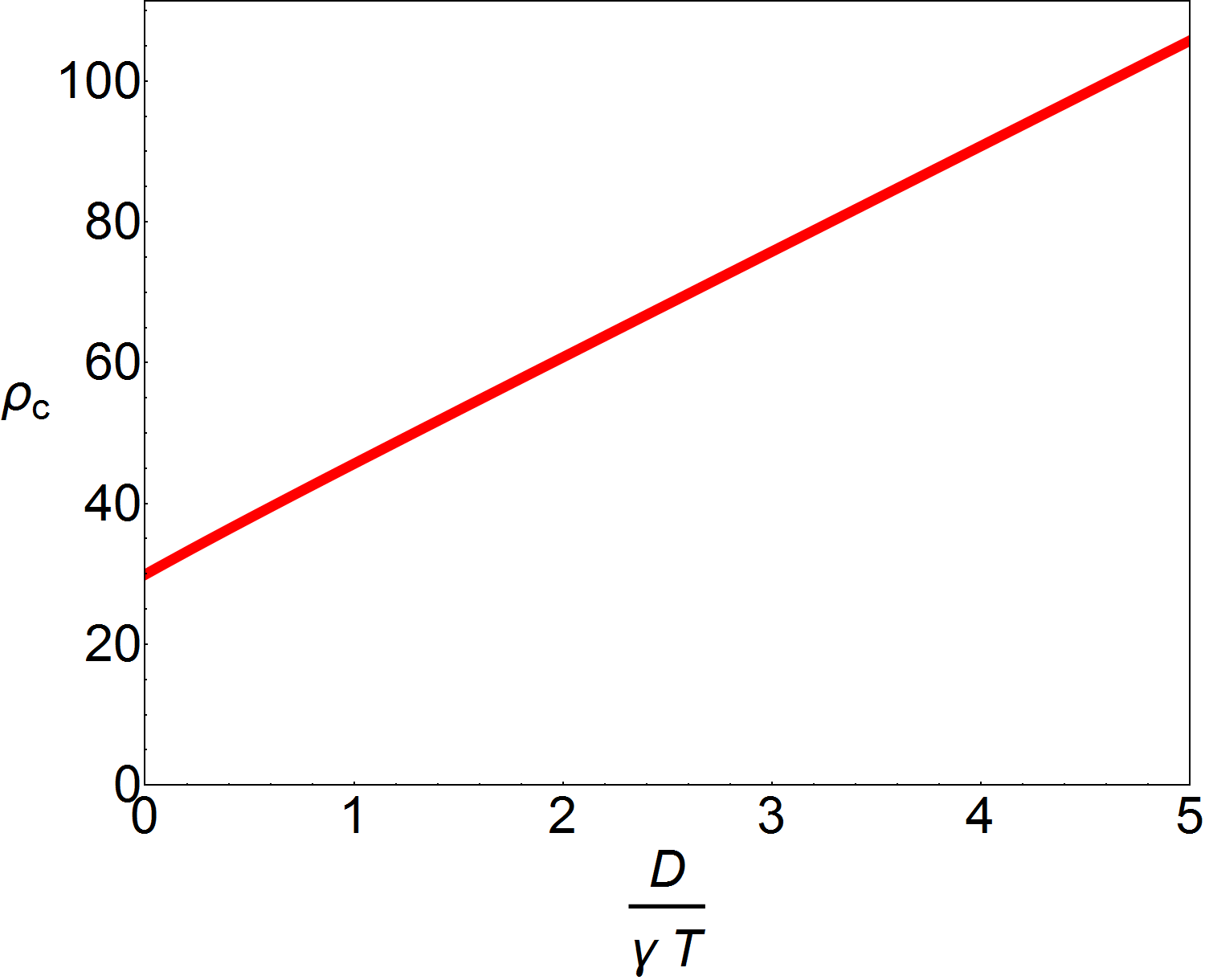}
\includegraphics[width=0.23\textwidth]{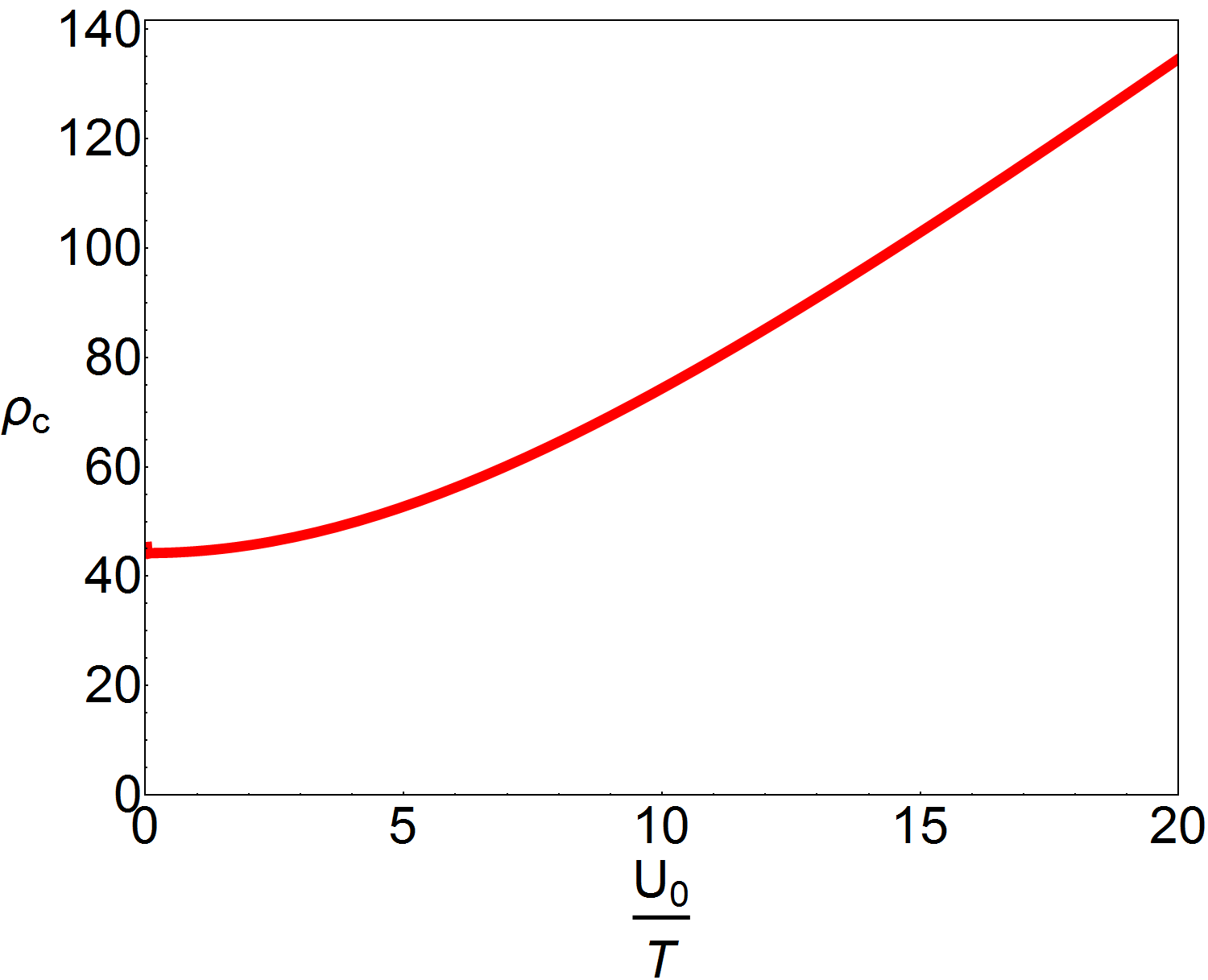}\\
\includegraphics[width=0.23\textwidth]{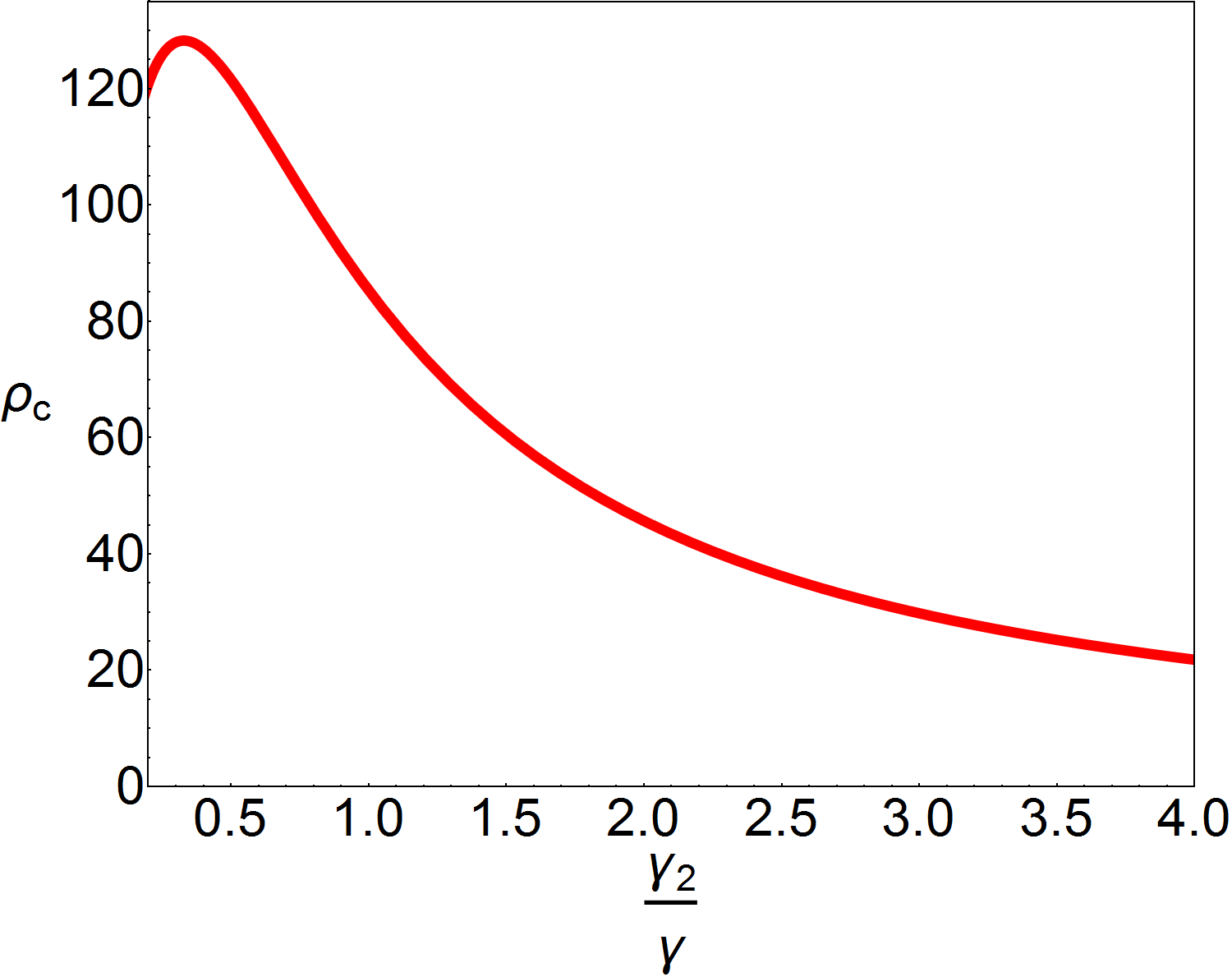}
\includegraphics[width=0.23\textwidth]{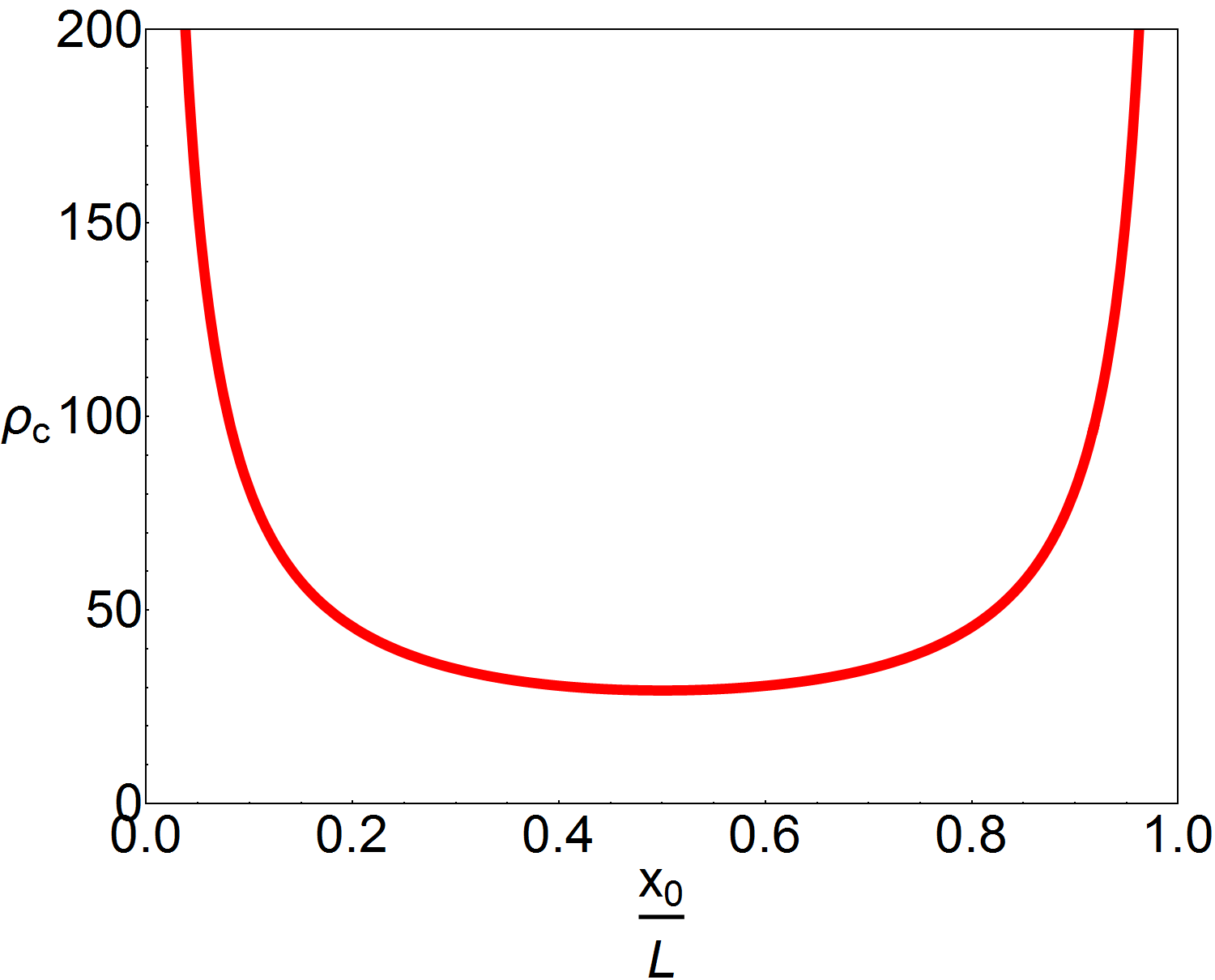}
\caption{The crossover value $\rho_c$ Eq.~\eqref{crossover} between slow and fast switching as a function of the noise intensity $D$ (top left), potential depth $U_0$ (top right), friction coefficient $\gamma_2$ (bottom left) and asymmetry parameter $x_0$ (bottom right). The respective remaining parameters are: $D = \gamma T, \ \gamma_2 = 2 \gamma, \ \gamma_1 = \gamma, \ U_0 = 2 T, \ x_0 = 0.2 L$. \label{fig:ratio} }
\end{figure}

As we discussed in the previous two sections, the current is proportional to $\rho = r \gamma L^2/T$ for small $\rho$, Eq.~\eqref{current-slow-result}, and inversely proportional to $\rho$ for large $\rho$, Eq.~\eqref{current-fast-result}.
For the piecewise linear potential Eq.~\eqref{piecewise-potential}, since the sign of the current is the same in both cases, it can be expected that the current exhibits a maximum in between the two limiting regimes.
An estimate for the position of this maximum is given by the crossover value $\rho_c$, defined as
\begin{align}
\rho_c \equiv \sqrt{\frac{J^{(1)}_{z,\text{fast}}}{J^{(1)}_{z,\text{slow}}}} \label{crossover},
\end{align}
where $J^{(1)}_{z,\text{slow}}$ ($J^{(1)}_{z,\text{fast}}$) denotes the first order correction to the dimensionless current for slow (fast) switching.
Equivalently, this corresponds to a crossover value in the switching rate $r_c = \rho_c/\tau_x$ with $\tau_x = \gamma L^2/T$.
The dimensionless quantity $\rho_c$ depends on the relative friction coefficients $\gamma_i/\gamma$, potential $U_0/T$, noise magnitude $D/(\gamma T)$ and position of the potential minimum $x_0/L$; it is shown as a function of these parameters in Fig.~\ref{fig:ratio}.
The crossover value $\rho$ increases linearly with the relative noise magnitude $D/(\gamma T)$, since larger $D$ corresponds to higher effective temperatures and thus faster position dynamics.
Arguing along the same lines, one would expect $\rho$ to decrease with increasing potential depth $U_0/T$, since a deeper potential should slow down the dynamics.
The observed behavior is the opposite; while both in the slow and fast switching regime the current decreases at large $U_0$, this decrease is actually faster in the slow switching regime.
This observation is, however, of limited value as for deep potentials, the relevant time scale for the position dynamics is no longer the time $\tau_x$ but instead the much larger escape time $\tau_\text{esc} \propto e^{U_0/T}$, so that we expect the system to be always in the fast switching regime for deep potentials.
Surprisingly, even if all the respective parameters are of order unity, $\rho_c$ is substantially larger than unity.
This is due to the first order coefficient of the current for fast switching being considerably larger than for slow switching.
This implies that the system carries larger current when the transitions between the internal states are fast compared to the position dynamics, than when the two timescales are similar.
We suspect that the cause for this at first glance surprising behavior is rooted in the close-to-equilibrium nature of the slow switching regime.
In this limit the system in state $i$ spends most of the time close to a quasi-equilibrium state with effective temperature $T_i$, and is only rarely driven out of equilibrium by a transition to a different state, which causes a current to flow.
By contrast for fast switching, even though the overall system can also be approximately described by an effective temperature (see Eq.~\eqref{p0-solution}), it is continuously driven out of equilibrium by the transitions between the internal states, leading to a larger overall current.

\begin{figure}[ht!]
\includegraphics[width=0.48\textwidth]{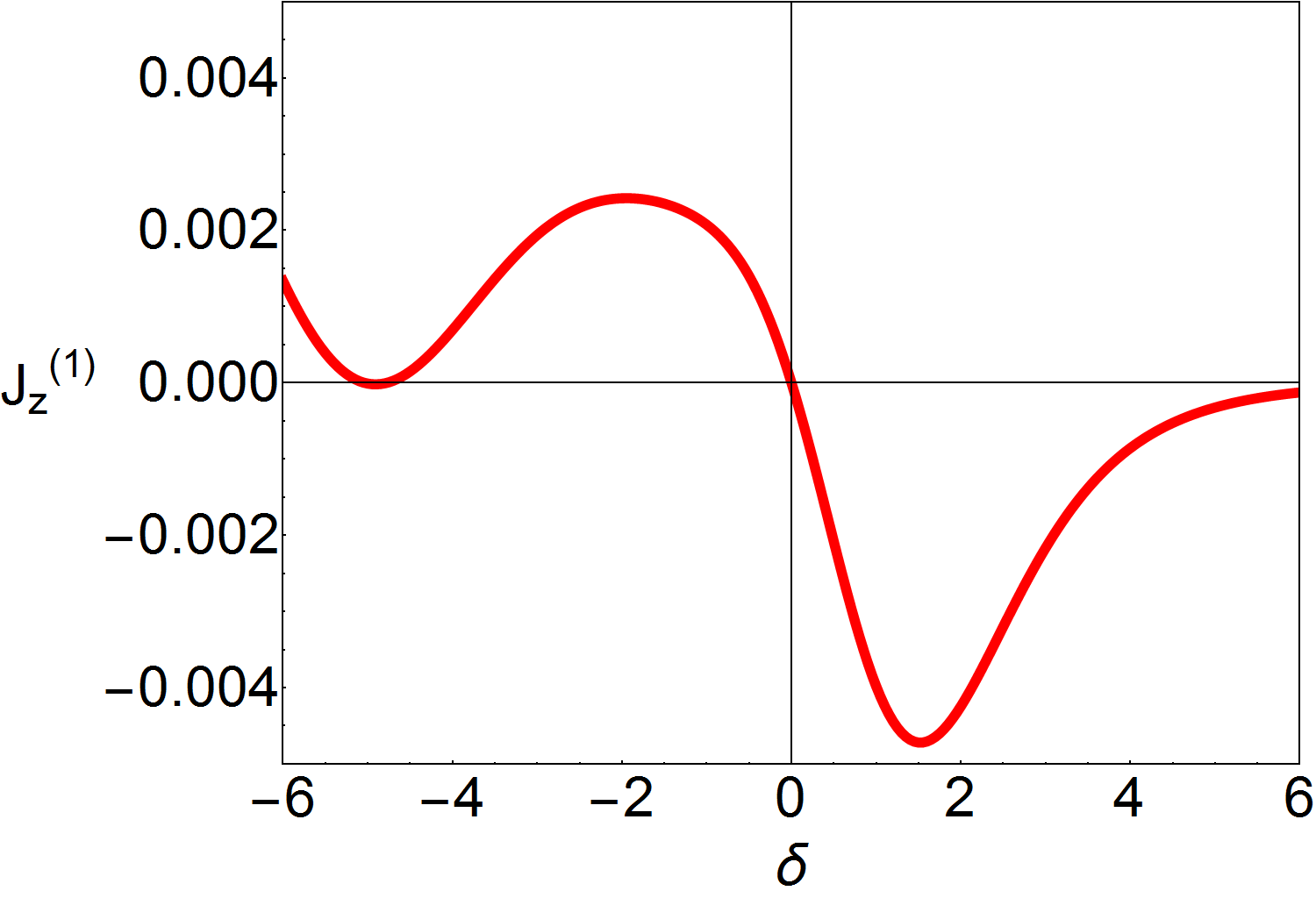}\\
\includegraphics[width=0.48\textwidth]{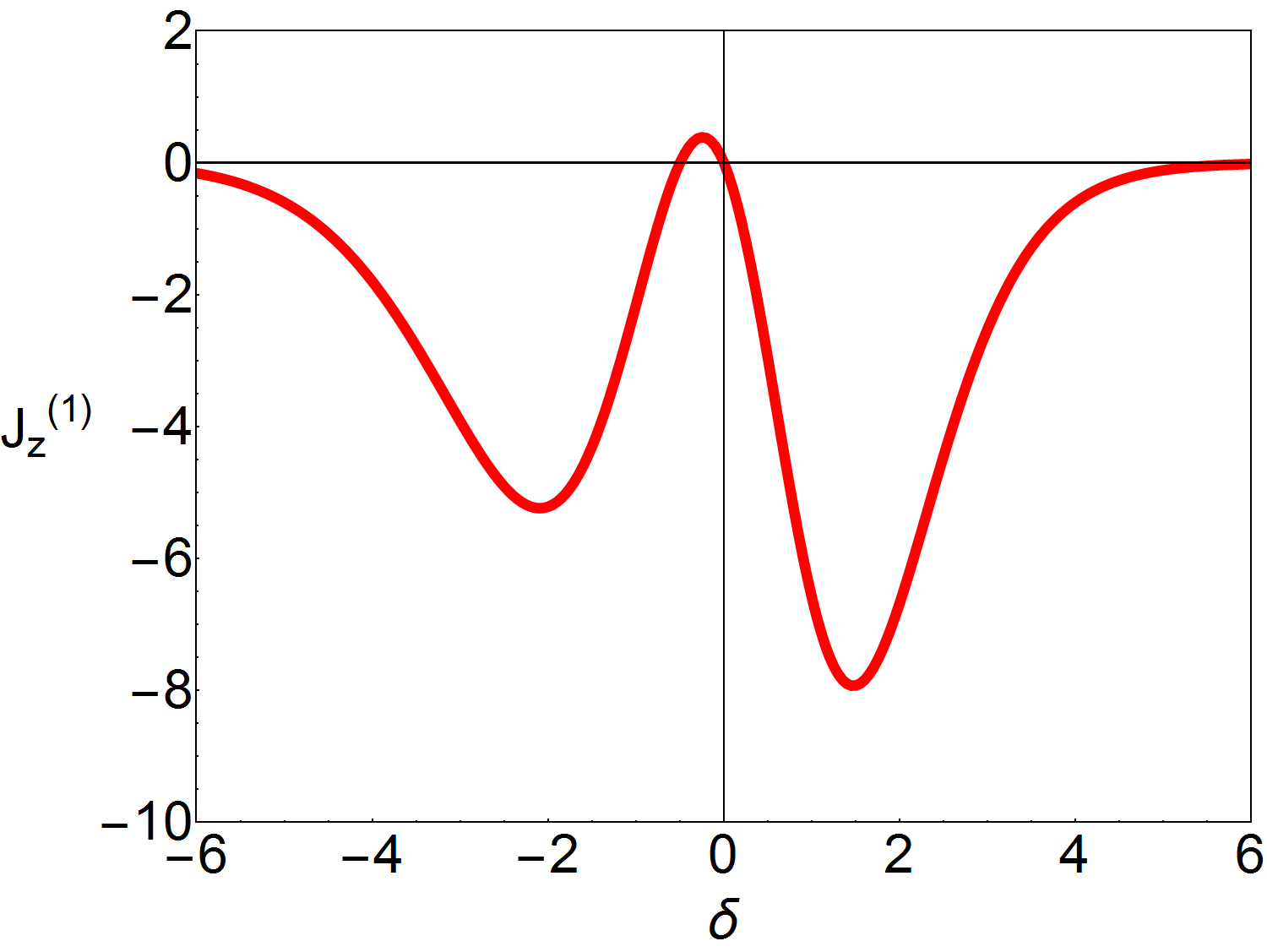}
\caption{The first order correction to the current for a potential $U(x) = U_0 [\sin(2 \pi x/L) + \delta (\sin(4 \pi x/L)/2 + \sin(6 \pi x/L)/3)]$, for slow (top) and fast (bottom) switching, shown as a function of the asymmetry parameter $\delta$. The remaining parameters are $D = \gamma T, \ \gamma_2 = 2 \gamma, \ \gamma_1 = \gamma, \ U_0 = 2 T$. Apart from being overall much larger, the fast switching current is also negative for $\delta \lesssim -1$, while the current for slow switching is positive. This implies a current reversal at some intermediate switching rate. \label{fig:reversal} }
\end{figure}
For the piecewise linear potential Eq.~\eqref{piecewise-potential}, the current always has the same direction in the slow and fast switching regimes. 
However, for other types of potential, this is not necessarily the case.
Choosing $U(x) = U_0 [\sin(2 \pi x/L) + \delta (\sin(4 \pi x/L)/2 + \sin(6 \pi x/L)/3)]$ and plotting the resulting first order correction for the current for fast and slow switching as a function of the asymmetry parameter $\delta$ (see Fig.~\ref{fig:reversal}), it is readily apparent that there exists a range of values for $\delta$ for which the current flows in opposite directions in the slow, respectively fast, switching regime.
This implies that there exists some intermediate switching rate $r$ at which the current changes sign and one thus observes a current reversal.
We note that also for this type of potential, the first order fast switching correction is about three orders of magnitude larger than the corresponding slow switching one, supporting the findings discussed above for the piecewise linear potential.

\subsection{Small noise} \label{sec-small-noise}
\begin{figure}[ht!]
\includegraphics[width=0.48\textwidth]{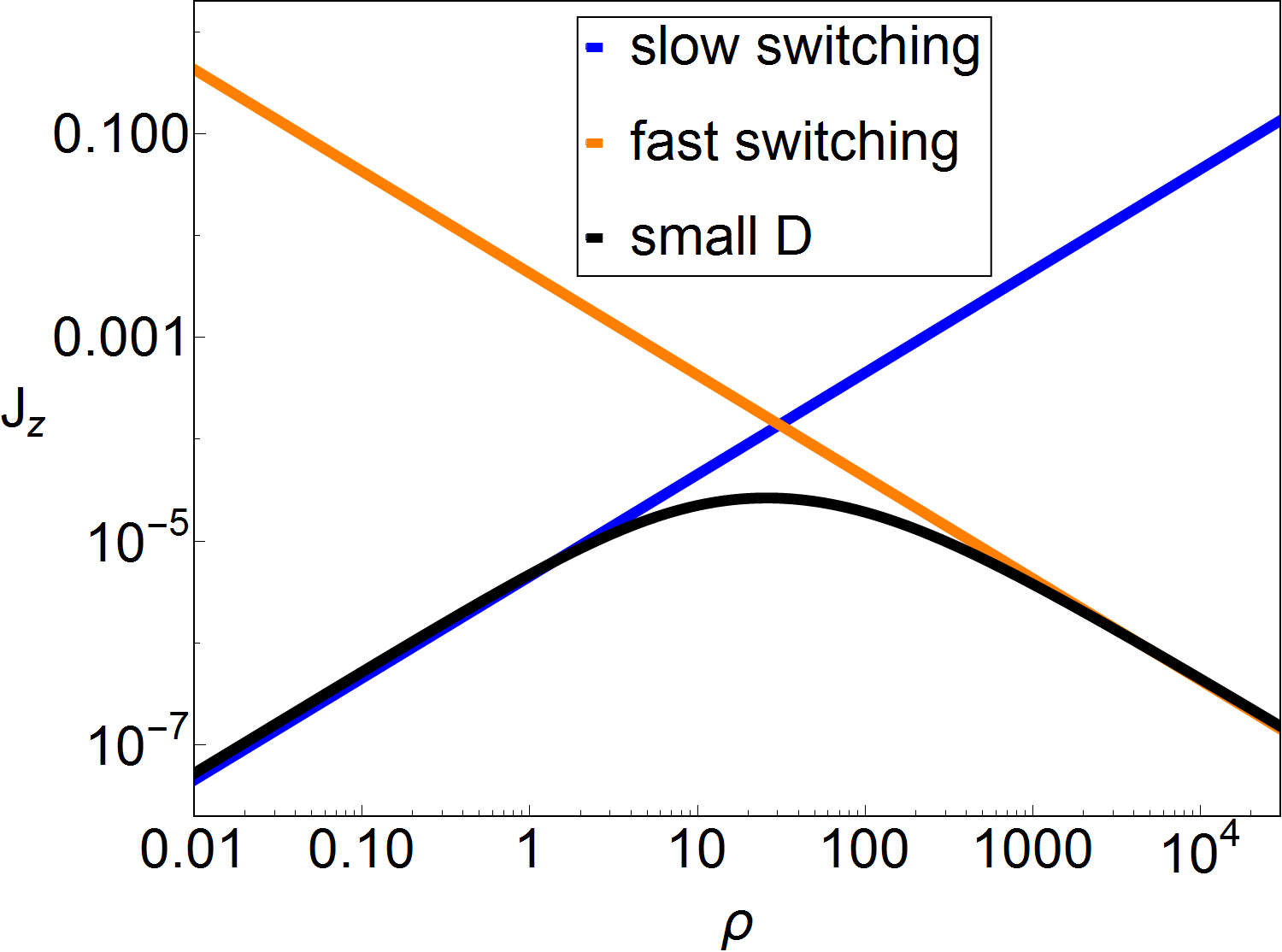}
\caption{The dimensionless current computed from the small $D$ (black), small $\rho$ (blue, Eq.~\eqref{current-slow-explicit}) and large $\rho$ (orange, Eq.~\eqref{current-fast-explicit}) expansions as a function of $\rho$ for $D = 0.05 \gamma T$. The remaining parameters are: $\gamma_2 = 2 \gamma, \ \gamma_1 = \gamma, \ U_0 = 2 T, \ x_0 = 0.2 L$. The result from the small $D$ expansion agrees very well with the slow and fast switching results in the respective limits. In between, as expected, it exhibits a maximum, whose position is approximately described by the crossover value $\rho_c \simeq 31$ (dashed line). \label{fig:current-small-d} }
\end{figure}

Since the current is driven by the added Gaussian white noise, we want to understand how the current scales in the limit where the noise intensity is small.
Expanding either Eq.~\eqref{current-slow-result} or Eq.~\eqref{current-fast-result} for small $D$, we find that the zeroth and first order contributions vanish, so that the total current is of order $D^2$, see also Fig.~\ref{fig:current-slow}.
While the zeroth order contribution has to vanish, as the system is in equilibrium for $D = 0$, the vanishing of the first order is non-trivial.
This observation is not restricted to the limits $r \ll T/(\gamma L^2)$ or $r \gg T/(\gamma L^2)$, but is in fact more general.
To see this, we consider an expansion of the probability density $P(x)$ and the current $J \simeq J^{(0)} + D J^{(1)} + D^2 J^{(2)}$ in terms of small $D \ll \gamma T$ similar to Eq.~\eqref{small-r-expansion}.
Plugging this into Eq.~\eqref{fokkerplanck}, it can be shown explicitly that the first order contribution to the current $J^{(1)}$ is always zero (see Appendix \ref{app-b}).
This agrees with the observation made in Ref.~\cite{Rei96}, that a ratchet driven by temperature differences seems to resist carrying a current.
The non-vanishing contribution to the current proportional to $D^2$ can be calculated explicitly only for simple cases of the potential $U(x)$.
For the piecewise linear potential Eq.~\eqref{piecewise-potential}, this procedure is carried out in Appendix \ref{app-b}.
The resulting current is compared to the slow and fast switching limits in Fig.~\ref{fig:current-small-d}.
Note that, as argued in Sec.~\ref{sec-crossover}, the maximum current occurs not for $\rho \simeq O(1)$ but at the considerably larger value $\rho = \rho_c \simeq 31$.

\section{Work extraction and efficiency} \label{sec-efficiency}
In the previous Section we derived analytic expressions for the current in several limits.
In particular, we can observe a non-vanishing current for any finite external noise intensity $D$.
This current can be used to extract work from the system and thus allows it to act as an engine driven by Gaussian white noise.
The simplest way to extract work is to apply a constant external load force $F_0$ opposite to the current.
As long as the external load is smaller than some stall value $F_\text{stall}$ the particle will on average move opposite to the load force and thus perform work on the load.
This can be quantified by the extracted power $P_\text{ex} = F_0 J_{x,\text{tot}} L$, where $J_{x,\text{tot}}$ is the total current in the presence of both load force and internal dynamics.
If the load force $F_0$ is small compared to the asymmetric potential $U(x)/L$, the total current decomposes into a contribution due to the external load, $J_{x,F} = F_0 c_J$, and the internal current $J_x$ that was calculated in Sec.~\ref{sec-limiting}.
Here $c_J$ depends on the details of the potential and the switching dynamics; we provide the corresponding expressions in App.~\ref{app-d} for slow (Eq.~\eqref{load-slow}) and fast (Eq.~\eqref{load-fast}) switching.
Since $J_x$ is to leading order independent of $F_0$, we can read of the stall force for which $J_{x,\text{tot}} = 0$,
\begin{align}
F_\text{stall} = - \frac{J_x}{c_J} .
\end{align}
The extracted power for small load is given by $P_\text{ext} = c_J L F_0^2 + J_x L F_0$.
The system performs work for $P_\text{ext} < 0$, which requires $J_x$ and $F_0$ to have opposite sign and $|F_0| < |F_\text{stall}|$.
The maximum power $P_\text{ext}^{*}$ is obtained for $F_0 = F_\text{stall}/2$
\begin{align}
P_\text{ext}^{*} = - \frac{1}{4} \frac{L}{c_J} J_x^2 .
\end{align}
Thus the maximum extractable power is proportional to the square of the internal current $J_x$, given by Eqs.~\eqref{current-slow-explicit} and \eqref{current-fast-explicit} for slow respectively fast switching.

Since our fluctuating friction ratchet permits the extraction of work and thus acts as an engine, it is natural to consider the efficiency with which the engine converts the external noise into work.
To define efficiency, we first need to specify the energy injected into the system as a consequence of the external noise.
From Eq.~\eqref{eq-fokkerplanck} and taking into account the external noise, we have for the change in the total energy of the particle
\begin{align}
\partial_t \langle U \rangle = \sum_i \frac{1}{\gamma_i} \bigg[ \langle U'' \rangle_i \Big(T + \frac{D}{\gamma_i} \Big) - \langle (U')^2 \rangle_i \bigg] .
\end{align}
Here, $\langle \ldots \rangle_i$ denotes an average with respect to $P_{i,t}(x)$, i.~e.~$\langle U \rangle_i$ is the energy of the fraction of particles in state $i$.
We identify the term proportional to $D$ as the energy injection rate due to the external noise.
In the steady state for two internal states and symmetric transition rates, the energy injection rate is given by
\begin{align}
\langle \dot{E}_D \rangle \simeq D \left\lbrace \begin{array}{ll}
\sum_{i = 1}^{2} \frac{\langle {U'}^2 \rangle_{T_i}}{2 T_i \gamma_i^2} + O(r) &\text{slow switching} \\[2 ex]
\frac{\langle {U'}^2 \rangle_{\overline{T}}}{2 \overline{T}} \frac{\gamma_1^2 + \gamma_2^2}{(\gamma_1 \gamma_2)^2} + O(r^{-1}) &\text{fast switching} \\[2 ex]
\frac{\langle {U'}^2 \rangle_{T}}{2 T} \frac{\gamma_1^2 + \gamma_2^2}{(\gamma_1 \gamma_2)^2} + O(D) &\text{small noise} ,
\end{array} \right.
\end{align}
where $\langle \ldots \rangle_T$ denotes an average over using a Boltzmann-Gibbs distribution at (effective) temperature $T$.
This expression is positive and to leading order independent of the switching rate $r$.
To define the efficiency, we have to compare this rate of energy injection to the extracted power $P_\text{ex}$
\begin{align}
\eta = -\frac{P_\text{ex}}{\langle \dot{E}_D \rangle} .
\end{align}
The maximum efficiency $\eta^{*}$ is thus obtained at maximum power $P^{*}_\text{ext}$.
For small external noise, the power is proportional to $D^4$ and thus the efficiency is proportional to $D^3$.
At moderate values of $D$ the efficiency is generally larger, however, here we only have asymptotic results for slow and fast switching.
Since the energy injection rate is independent of the switching rate $r$, but the power is proportional to $r^2$ for slow switching, respectively $r^{-2}$ for fast switching, the efficiency is proportional to $r^2$, respectively $r^{-2}$.
Thus, at least in the limits discussed above, the efficiency of the ratchet, like the current, is asymptotically small; only a small proportion of the injected energy is converted into useful work.
However, this neglects the fact that Gaussian white noise is usually a readily available and often naturally occurring resource, which might make the ratchet viable as an engine despite its low efficiency.



\section{Discussion} \label{sec-discussion}
With an internal degree of freedom with a state-dependent friction coefficient, an overdamped Brownian particle can be driven out of equilibrium by the addition of state-independent Gaussian white noise.
As we have shown, this can be utilized to generate a current and extract work from the uncorrelated Gaussian fluctuations of the noise.
Previous studies have focused on ratchets driven by correlated \cite{Mag93,Mil94,Doe94,Bar96} or non-Gaussian noise \cite{Luc95,Luc97,Cze97}, which require a specially tailored noise source.
Gaussian white noise, on the other hand, occurs naturally as the coarse-graining limit of any weakly correlated, finite-variance noise and thus occurs commonly in the description of many physical systems, often as an equilibrium noise.
Our analysis underlines that the difference between equilibrium and nonequilibrium noise is not only due to the noise itself but always has to be determined in the context of the system.
This means that even uncorrelated Gaussian noise, which is an ideal realization of \enquote{true} randomness, can serve as a resource from which work can be extracted, given a suitably designed system.
Thus Gaussian white noise can potentially serve as a resource to power microscopic engines.

\begin{acknowledgments}
\textbf{Acknowledgments.} The present study was supported by KAKENHI (Nos. 25103002 and 26610115).
A.~D.~was employed as an International Research Fellow of the Japan Society for the Promotion of Science.
\end{acknowledgments}

\appendix

\section{Current for sawtooth potential} \label{app-a}

For the piecewise linear sawtooth potential Eq.~\eqref{piecewise-potential}, the first order slow switching contribution to the current Eq.~\eqref{current-slow-result} can be evaluated explicitly.
We restrict ourselves to two internal states $\lbrace 1, 2 \rbrace$ and assume symmetric transition rates $\alpha_{1 2} = \alpha_{2 1} = 1$.
The result reads
\begin{widetext}
\begin{align}
J_x^{(1)} = \frac{\big(1 - 2 \frac{x_0}{L} \big) }{16 \sinh^2\big(\frac{\beta_1}{2}\big) \sinh^2\big(\frac{\beta_2}{2}\big) } \Bigg[ \beta_1 \cosh(\beta_2) + \beta_2 \cosh(\beta_1) + \frac{\beta_1 + \beta_2}{\beta_2 - \beta_1} \Big( \sinh(\beta_1) + \beta_1 - \sinh(\beta_1 - \beta_2) - \sinh(\beta_2) - \beta_2 \Big) \Bigg]. \label{current-slow-explicit}
\end{align}
\end{widetext}
Here we defined $\beta_i = U_0/T_i$ with the effective temperatures $T_i = T + D/\gamma_i$.
As expected, the current vanishes for $x_0 = L/2$ or $T_1 = T_2$ (equivalent to $D=0$ or $\gamma_1 = \gamma_2$).

\section{Small $D$ for arbitrary $r$} \label{app-b}
For small $D \ll \gamma T$ and arbitrary $r$, we expand the probability density similar to Eq.~\eqref{small-r-expansion}, $P_i(x) \simeq P_i^{(0)}(x) + D P_i^{(1)}(x) + O(D^2)$.
We then have from Eq.~\eqref{fokkerplanck} to first order in $D$
\begin{align}
&\frac{1}{\gamma_i} \partial_x \Big[ U'(x) + T \partial_x \Big] P_i^{(1)}(x) + \frac{1}{\gamma_i^2} \partial_x^2 P_i^{(0)}(x) \nonumber \\
\quad &= \sum_{j \neq i} \big( r_{j i} P_i^{(1)}(x) - r_{i j} P_j^{(1)}(x) \big)  \label{multiple-small-D} \\
&\text{with} \quad P_i^{(0)}(x) = \frac{\mathcal{P}_i}{Z} e^{-\frac{U(x)}{T}}, \quad Z = \int_0^L \text{d}x \ e^{-\frac{U(x)}{T}} \nonumber .
\end{align}
Taking the sum over the $N$ equations, the term involving the transition rates cancels and we are left with
\begin{align}
&\sum_i\bigg[\frac{1}{\gamma_i} \partial_x \Big[ U'(x) + T \partial_x \Big] P_i^{(1)}(x) + \frac{1}{\gamma_i^2} \partial_x^2 P_i^{(0)}(x)\bigg] = 0 .
\end{align}
Due to the definition of the current Eq.~\eqref{current-def}, we obtain upon integration
\begin{align}
&\Big[ U'(x) + T \partial_x \Big] \widetilde{P}^{(1)}(x) + \sum_i \frac{1}{\gamma_i^2} \partial_x P_i^{(0)}(x) = -J_x^{(1)} ,
\end{align}
where we defined
\begin{align}
\widetilde{P}^{(1)}(x) = \sum_i \frac{1}{\gamma_i} P_i^{(1)}(x) \label{p-tilde}.
\end{align}
This is readily solved for $\widetilde{P}^{(1)}(x)$,
\begin{align}
\widetilde{P}^{(1)}(x) &= \frac{1}{T} e^{-\frac{U(x)}{T}} \bigg[ \tilde{N} \label{p-tilde-sol} \\
& -\int_0^x \text{d}y \ e^{\frac{U(y)}{T}} \Big[ \sum_i \frac{1}{\gamma_i^2} \partial_y P_i^{(0)}(y) + J_x^{(1)} \Big]  \bigg] \nonumber .
\end{align}
Since every $P_i^{(1)}(x)$ has to be periodic, the same is true for $\widetilde{P}^{(1)}(x)$ and we find
\begin{align}
J_x^{(1)} = -\frac{1}{Z^+} \sum_i \frac{1}{\gamma_i^2} \int_0^L \text{d}y \ e^{\frac{U(y)}{T}} \partial_y P_i^{(0)}(y) ,
\end{align}
with $Z^+ = \int_0^L \text{d}x \ e^{U(x)/T}$.
Using the expression for $P_i^{(0)}$ from Eq.~\eqref{multiple-small-D}, it is easy to see that the integral yields zero and the first order contribution to the current for small $D$ vanishes.
For the second order contribution, we find in complete analogy
\begin{align}
J_x^{(2)} = -\frac{1}{Z^+} \sum_i \frac{1}{\gamma_i^2} \int_0^L \text{d}y \ e^{\frac{U(y)}{T}} \partial_y P_i^{(1)}(y) ,
\end{align}
where the $P_i^{(1)}$ are determined by the solution of Eq.~\eqref{multiple-small-D}.
For general $U(x)$ there exists no closed form solution for $P_i^{(1)}$, however, we can solve the equations in special cases.
First we note that, from Eq.~\eqref{p-tilde-sol} using the normalization condition $\int_0^L \text{d}x \ \widetilde{P}^{(1)}(x) = 0$
\begin{align}
\widetilde{P}^{(1)}(x) = \frac{1}{Z T^2} \sum_i \frac{\mathcal{P}_i}{\gamma_i^2} \big( U(x) - \langle U \rangle_0  \big) e^{-\frac{U(x)}{T}} .
\end{align}
Comparing this to the definition of $\widetilde{P}^{(1)}$, Eq.~\eqref{p-tilde}, we define
\begin{align}
P_i^{(1)}(x) = \frac{1}{Z T^2} \frac{\mathcal{P}_i}{\gamma_i} \big( U(x) - \langle U \rangle_0  \big) e^{-\frac{U(x)}{T}} + g_i(x).
\end{align}
The first part on the right hand side is the trivial correction due to finite $D$, which due to an increased temperature $T \rightarrow T + D/\gamma_i$ increases the probability for the particle to be at larger values of the potential compared to the average $\langle U \rangle_0$ at temperature $T$.
It is easy to verify by direct computation that this trivial term does not contribute to the current. 
The functions $g_i$ then satisfy the equations
\begin{align}
\frac{1}{\gamma_i} &\partial_x \Big[ U'(x) + T \partial_x \Big] g_i(x) = \sum_j \big( r_{j i} g_i(x) - r_{i j} g_j(x) \big) \nonumber \\
& \qquad \qquad + \frac{\xi_i}{Z T^2} \big( U(x) - \langle U \rangle_0  \big) e^{-\frac{U(x)}{T}},
\end{align}
with $\xi_i = \sum_j (r_{j i} \mathcal{P}_i/\gamma_i - r_{i j} \mathcal{P}_j / \gamma_j)$ and the additional condition due to Eq.~\eqref{p-tilde}
\begin{align}
\sum_i \frac{1}{\gamma_i} g_i(x) = 0 \label{gi-condition}.
\end{align}
In terms of the $g_i$, the second order correction to the current reads
\begin{align}
J_x^{(2)} = -\frac{1}{Z^+} \sum_i \frac{1}{\gamma_i^2} \int_0^L \text{d}y \ e^{\frac{U(y)}{T}} \partial_y g_i(y) \label{current-small-d} .
\end{align}
For two internal states with symmetric transition rates $r_{1 2} = r_{2 1} = r$, we can use Eq.~\eqref{gi-condition} to eliminate $g_2$ and obtain
\begin{align}
&\partial_x \Big[ U'(x) + T \partial_x \Big] g_1(x) = r (\gamma_1 + \gamma_2) g_1(x)  \nonumber \\
& \qquad \qquad + \frac{r}{2 Z T^2} \Big(1 - \frac{\gamma_1}{\gamma_2} \Big) \big( U(x) - \langle U \rangle_0  \big) e^{-\frac{U(x)}{T}} \label{gi-equation}.
\end{align}
This equation can be solved explicitly for the piecewise linear potential Eq.~\eqref{piecewise-potential}, where $U'(x) = \alpha = \text{const}$.
In this case, the general solution to Eq.~\eqref{gi-equation} reads
\begin{align}
g_i(x) = \frac{1}{\epsilon} \Bigg[ &e^{-\frac{(\alpha + \epsilon) x}{2 T}} \bigg[c_1 - \int \text{d}x \ e^{\frac{(\alpha + \epsilon) x}{2 T}} h_1(x) \bigg] \\
& + e^{-\frac{(\alpha - \epsilon) x}{2 T}} \bigg[c_2 + \int \text{d}x \ e^{\frac{(\alpha - \epsilon) x}{2 T}} h_1(x) \bigg] \Bigg] \nonumber ,
\end{align}
where we defined
\begin{subequations}
\begin{align}
\epsilon &= \sqrt{\alpha^2 + 4 r T (\gamma_1 + \gamma_2)} \\
h_1(x) &= \frac{r}{2 Z T^2} \Big(1 - \frac{\gamma_1}{\gamma_2} \Big) \big( U(x) - \langle U \rangle_0  \big) e^{-\frac{U(x)}{T}} .
\end{align}%
\end{subequations}
For the potential Eq.~\eqref{piecewise-potential} we have $\alpha^l = U_0/x_0$ for $0 < x < x_0$ and $\alpha^r = - U_0/(L-x_0)$ for $x_0 < x < L$.
Matching $g_1$ and its derivative at the boundaries,
\begin{align}
g_1^r (x_0) &= g_1^l(x_0), \qquad g_1^l(0) = g_1^r(L), \\
{g_1^{r}}' (x_0) &= {g_1^{l}}'(x_0) + \frac{U_0}{T} g_1^{r}(x_0) \frac{L}{x_0(L-x_0)}, \nonumber \\
{g_1^{l}}' (0) &= {g_1^{r}}'(L) - \frac{U_0}{T} g_1^{l}(0) \frac{L}{x_0(L-x_0)}, \nonumber
\end{align}
determines the four coefficients $c_{1,2}^{l,r}$.
Plugging the results for $g_1$ and $g_2$ into Eq.~\eqref{current-small-d} then yields the desired expression for the current.
This above procedure is straightforward but tedious and we use Mathematica to solve and evaluate the resulting expressions.

\section{Current due to external load} \label{app-d}
Taking into account a constant external load force $F_0$, Eq.~\eqref{fokkerplanck-dimensionless} changes to
\begin{align}
\frac{1}{\nu_i} \partial_z \Big[u'(z) - f &+ \theta_i \partial_z \Big] p_i(z) = \rho \sum_{j} \big(\alpha_{j i} p_i - \alpha_{i j} p_j \big) \label{fokkerplanck-dimensionless-load} ,
\end{align}
where $f = F_0 L / T$.
For small load $f \ll 1$, we can expand $p_i(z) \simeq p_i^{(0)} + f p_i^{(1)}$.
Here $p_i^{(0)}$ is the result for zero load force, which was discussed in Sec.~\ref{sec-limiting}.
Integrating over $z$, we have
\begin{align}
\frac{1}{\nu_i} &\Big[u'(z) + \theta_i \partial_z \Big] p^{(1)}_i(z) \\
& = \rho \sum_{j} \int_{0}^{z} \text{d}y \ \big(\alpha_{j i} p^{(1)}_i(y) - \alpha_{i j} p^{(1)}_j(y) \big) \nonumber \\
&\qquad + \frac{1}{\nu_i} p_i^{(0)}(z) - J_i^{(1)} \nonumber  ,
\end{align}
where the constants $J_i^{(1)}$ are related to the total current via $J_z \simeq \sum_i (J_i^{(0)} + f J_i^{(1)})$.
For slow switching, we can now further expand for small $\rho \ll 1$, $p_i^{(k)} \simeq p_i^{(k,0)} + \rho p_i^{(k,1)}$.
Since the term $p_i^{(1,1)}$ leads to a contribution of order $f \rho$, we neglect it and only consider the leading order term, for which we have
\begin{align}
\frac{1}{\nu_i} &\Big[u(z) + \theta_i \partial_z \Big] p^{(1,0)}_i(z) = \frac{1}{\nu_i} p_i^{(0,0)}(z) - J_i^{(1)} .
\end{align}
The function $p_i^{(0,0)}$ is just the effective Boltzmann-Gibbs distribution Eq.~\eqref{r0-solution}.
Solving for $p^{(1,0)}_i$ and demanding periodicity, we obtain
\begin{align}
J_i^{(1)} = \frac{\mathcal{P}_i}{\nu_i Z_i Z_i^+},
\end{align}
where $Z_i$ and $Z_i^+$ were defined in Sec.~\ref{sec-slow}.
The total current is then given by
\begin{align}
J_{z,\text{tot}} \simeq \sum_i \frac{\mathcal{P}_i}{\nu_i Z_i Z_i^+} f + J_z + O(\rho f) ,
\end{align}
where $J_z$ is given by Eq.~\eqref{current-slow-result} and is proportional to $\rho$ and independent of $f$.
In terms of dimensionful quantities we have $J_{x,\text{tot}} = J_{z,\text{tot}}/\tau_x $ and thus
\begin{align}
J_{x,\text{tot}} \simeq \frac{F_0}{L} \sum_i \frac{\mathcal{P}_i}{\gamma_i Z_i Z_i^+}  + J_x . \label{load-slow}
\end{align}
Following an analogous calculation using the results of Sec.~\ref{sec-fast}, we find in the fast switching regime
\begin{align}
J_{x,\text{tot}} \simeq \frac{F_0}{L} \frac{1}{2 \overline{\gamma} \overline{Z} \overline{Z}^+}  + J_x . \label{load-fast}
\end{align}

\bibliography{bib}

\begin{thebibliography}{27}
\bibitem [{not()}]{note1}%
  \BibitemOpen
  \href@noop {} {}\bibinfo {note} {Since the zeroth order contribution already
  satisfies normalization $\int_0^L \text{d}x \ P_i(x) = \int_0^L \text{d}x \
  P_i^{(0)}(x) = \mathcal{P}_i$, all sub-leading contributions need to be
  normalized to zero $\int_0^L \text{d}x \ P_i^k(x) = 0$ for $k \geq
  1$.}\BibitemShut {Stop}%
\end{thebibliography}

\end{document}